\documentclass[longauth]{aa} 

\newcommand{\micron}{$\mu$m}
\newcommand{\paa}{Pa$\alpha$}
\newcommand{\pab}{Pa$\beta$}
\newcommand{\ha}{H$\alpha$}
\newcommand{\kms}{km~s$^{-1}$}
\newcommand{\msunperyear}{M$_{\odot}$\,yr$^{-1}$}
\newcommand{\msun}{M$_{\odot}$}

\newcommand{\sfrpaa}{SFR$_{\mathrm{Pa\alpha}}$}
\newcommand{\sfrpab}{SFR$_{\mathrm{Pa\beta}}$}
\newcommand{\sfrir}{SFR$_{\mathrm{IR}}$}
\newcommand{\sfruv}{SFR$_{\mathrm{UV}}$}

\usepackage{graphicx}
\usepackage{txfonts}
\usepackage[colorlinks=true,linkcolor=red,anchorcolor=blue,citecolor=blue,filecolor=black,menucolor=black,runcolor=black,urlcolor=blue]{hyperref}
\begin{document}

   \title{Clumpy star formation and an obscured nuclear starburst  in the luminous dusty z=4 galaxy GN20 seen by MIRI/JWST}

\titlerunning{Clumpy star formation and a extremely obscured nuclear starburst in GN20}


   \author{A. Bik
          \inst{\ref{inst:Stockholm}}
 \and J. \'Alvarez-M\'arquez\inst{\ref{inst:CAB}}  \and L. Colina\inst{\ref{inst:CAB}} \and A. Crespo G\'omez\inst{\ref{inst:CAB}} \and  F. Peissker\inst{\ref{inst:Koln}}  \and F. Walter\inst{\ref{inst:MPIA}} \and L. A. Boogaard\inst{\ref{inst:MPIA}}   \and G. {\"O}stlin\inst{\ref{inst:Stockholm}}
\and T.R. Greve\inst{\ref{inst:DTU}, \ref{inst:DAWN}, \ref{inst:UCL}} 
 \and G. Wright\inst{\ref{inst:UKATC}} 
\and A. Alonso-Herrero\inst{\ref{inst:CAB-ESAC}}    
\and K.I. Caputi\inst{\ref{inst:Groningen},\ref{inst:DAWN}} 
\and L. Costantin\inst{\ref{inst:CAB}}  
\and A. Eckart\inst{\ref{inst:Koln}} 
\and S. Gillman\inst{\ref{inst:DTU},\ref{inst:DAWN}} 
\and J. Hjorth\inst{\ref{inst:DARK}} 
\and E. Iani\inst{\ref{inst:Groningen}}
\and I. Jermann\inst{\ref{inst:DTU}, \ref{inst:DAWN}} 
\and A. Labiano\inst{\ref{inst:Telespazio},\ref{inst:CAB}}
\and D. Langeroodi\inst{\ref{inst:DARK}} 
\and J. Melinder\inst{\ref{inst:Stockholm}}
\and P. G. P\'erez-Gonz\'alez\inst{\ref{inst:CAB}} 
\and J.P. Pye\inst{\ref{inst:Leicester}} 
\and P. Rinaldi\inst{\ref{inst:Groningen}} 
\and T. Tikkanen\inst{\ref{inst:Leicester}} 
   \and P. van der Werf\inst{\ref{inst:Leiden}} 
   \and M. G\"udel\inst{\ref{inst:Vienna}, \ref{inst:MPIA}, \ref{inst:ETH}} 
\and Th. Henning\inst{\ref{inst:MPIA}}
\and P.O. Lagage\inst{\ref{inst:AIM}}
\and T. Ray\inst{\ref{inst:Dublin}}
   \and E.F. van Dishoeck\inst{\ref{inst:Leiden}}  
     }




 %


   \institute{
   Department of Astronomy, Stockholm University, Oscar Klein Centre, AlbaNova University Centre, 106 91 Stockholm, Sweden\\  \email{arjan.bik@astro.su.se} \label{inst:Stockholm}
   \and Centro de Astrobiolog\'{\i}a (CAB), CSIC-INTA, Ctra. de Ajalvir km 4, Torrej\'on de Ardoz, E-28850, Madrid, Spain  \label{inst:CAB}
    \and I.Physikalisches Institut der Universit\"at zu K\"oln, Z\"ulpicher Str. 77, 50937 K\"oln, Germany \label{inst:Koln}
    \and Max-Planck-Institut f\"ur Astronomie, K\"onigstuhl 17, 69117 Heidelberg, Germany\label{inst:MPIA}
    \and DTU Space, Technical University of Denmark, Elektrovej 327, 2800 Kgs. Lyngby, Denmark \label{inst:DTU}
    \and Cosmic Dawn Centre (DAWN), Copenhagen, Denmark \label{inst:DAWN}
    \and Department of Physics and Astronomy, University College London, Gower Place, London WC1E 6BT, UK \label{inst:UCL}
    \and UK Astronomy Technology Centre, Royal Observatory Edinburgh, Blackford Hill, Edinburgh EH9 3HJ, UK \label{inst:UKATC} 
   \and Centro de Astrobiolog\'ia (CAB), CSIC-INTA, Camino Viejo del Castillo s/n, 28692 Villanueva de la Ca\~{n}ada, Madrid, Spain \label{inst:CAB-ESAC}  
    \and Kapteyn Astronomical Institute, University of Groningen, P.O. Box 800, 9700 AV Groningen, The Netherlands \label{inst:Groningen}   
   \and DARK, Niels Bohr Institute, University of Copenhagen, Jagtvej 128, 2200 Copenhagen, Denmark \label{inst:DARK}    
    \and Telespazio UK for the European Space Agency, ESAC, Camino Bajo del Castillo s/n, 28692 Villanueva de la Ca\~{n}ada, Spain \label{inst:Telespazio}
    \and School of Physics \& Astronomy, Space Research Centre, Space Park Leicester, University of Leicester, 92 Corporation Road, Leicester LE4 5SP, UK \label{inst:Leicester} 
   \and Leiden Observatory, Leiden University, PO Box 9513, 2300 RA Leiden, The Netherlands \label{inst:Leiden}              
    \and University of Vienna, Department of Astrophysics, Türkenschanzstrasse 17, 1180 Vienna, Austria \label{inst:Vienna}
    \and Institute of Particle Physics and Astrophysics, ETH Zurich, Wolfgang-Pauli-Str 27, 8093 Zurich, Switzerland \label{inst:ETH} 
    \and Universit\'e Paris-Saclay, Universit\'e Paris Cit\'e, CEA, CNRS, AIM, F-91191 Gif-sur-Yvette, France \label{inst:AIM}
   \and Dublin Institute for Advanced Studies, Astronomy \& Astrophysics Section, 31 Fitzwilliam Place, Dublin, D02 XF86 Ireland \label{inst:Dublin}
    }

   \date{}

 
  \abstract{Dusty star-forming galaxies emit most of their light at far-IR to mm wavelengths as their star formation is highly obscured. Far-IR and mm observations have revealed their dust, neutral and molecular gas properties.  
  The sensitivity of JWST at rest-frame optical and near-infrared wavelengths now allows the study of the stellar and ionized gas content. We investigate the spatially resolved distribution and kinematics of the ionized gas in GN20, a dusty star forming galaxy at $z$=4.0548. We present deep MIRI/MRS integral field spectroscopy of the near-infrared rest-frame emission of GN20. We detect spatially resolved \paa, out to a radius of 6 kpc, distributed in a clumpy morphology. The star formation rate derived from \paa\ (144 $\pm$ 9 \msunperyear) is only 7.7 $\pm 0.5 $\% of the infrared star formation rate (1860 $\pm$ 90 \msunperyear).  We attribute this to very high extinction (A$_V$ = 17.2 $\pm$ 0.4 mag, or A$_{V,mixed}$ = 44 $\pm$ 3 mag), especially in the nucleus of GN20, where only faint \paa\ is detected, suggesting a deeply buried starburst. We identify four, spatially unresolved, clumps in the \paa\ emission. Based on the double peaked \paa\ profile we find that each clump consists of at least two sub-clumps. We find mass upper limits consistent with them being formed in a gravitationally unstable gaseous disk. The UV bright region of GN20 does not have any detected \paa\ emission, suggesting an age of more than 10 Myrs for this region of the galaxy.
  From the rotation profile of \paa\ we conclude that the gas kinematics are rotationally dominated and the  $v_{rot}/\sigma_{m} = 3.8 \pm 1.4$ is similar to low-redshift LIRGs. From the \paa\ kinematics we cannot  distinguish between a rotational profile of a large disk and a late stage merger mimicking a disk. We speculate that GN20 is in the late stage of a major merger, where the clumps in a large gas rich disk are created by the major merger, while the central starburst is driven by the merger event.}


   \maketitle

\section{Introduction}

Dusty Star Forming Galaxies (DSFGs) are extreme infrared luminous galaxies with short intense starburst episodes, forming stars at rates from 500 to several 1000 \msunperyear \citep[see ][for a review]{Casey14}. They are the most luminous starbursts in the Universe and are considered  the progenitors of the massive quiescent galaxies between $z$\,$\sim$\,2\,$-$\,3 \citep{Valentino20}. These galaxies emit most of their light at  (far) infrared to radio wavelengths and are optically obscured due to a significant dust content which attenuates the light. They are identified as sub-millimeter galaxies (SMGs) in deep sub-millimeter surveys \citep[e.g.][]{Hughes98,Borys03,Pope06,Vieira13} and represent an important phase of cosmic star formation missed by the deep rest-frame optical and ultraviolet surveys \citep{Bouwens21,Bouwens22,Finkelstein23,Perez23}. More recent surveys with ALMA have concluded that the infrared bright galaxies dominate the cosmic star formation rate below $z$\,$\sim$\,4 \citep{Perez05,Gruppioni13,Zavala21}, while at higher redshift their contribution goes down to 10-30\% at $z$\,$\sim$\,6\,$-$\,7 \citep{Gruppioni20,Algera23,Barrufet23Rebels}.

There are several scenarios presented in the literature for the triggering mechanism of the extreme starburst in DSFGs. Many DSFGs are thought to be starburst dominated major mergers. ALMA observations reveal the disturbed morphology and kinematics in several of those galaxies \citep[e.g.][]{Gomez18,Riechers20,Ginolfi20,Spilker22,Alvarez23}. Several of these galaxies are located in over densities or are members of proto clusters, making interactions a likely cause for their extreme starburst \citep[e.g.][]{Daddi09,Drake20}. Their merger origin would make the DSFGs high redshift analogues of ultra luminous infrared galaxies (ULIRGs) in the local universe. 

The other scenario is driven by the observations of star-forming galaxies at high redshift showing gas rich disks \citep[e.g.][]{Genzel08,Elmegreen09}. In this scenario the star formation is fuelled by cold mode accretion \citep[e.g.][]{DekelNature09,Dekel09}, where the high star formation in the gas rich disk is maintained by smooth infall and accretion of gas. Due to the high gas densities, the disk becomes unstable due to gravitational instabilities \citep{Krumholz18}, resulting in the formation of giant star forming clumps in the disk \citep{Dekel09,Romeo10,Ceverino10,Mandelker14,Romeo14} with  kpc sizes and masses of $\sim$10$^9$ \msun\ \citep{Genzel08,Elmegreen09,Ceverino10}. These clumps could then migrate further into the center of the galaxy due to dynamical friction over timescales of several hundred Myr \citep[e.g.][]{Mandelker14} and contribute to the formation of the thick disk and the bulge growth \citep{Elmegreen08,Elmegreen09b,Ceverino10}. Observations of individual galaxies support this scenario by finding trends between galactocentric distance and age of the clumps \citep{Adamo13,Cava18}.

Due to the fact that the DSFGs have high extinction, almost all we know about these galaxies is derived from their molecular gas and  dust content \citep[e.g.][]{Hodge12,Hodge15} tracing mostly the neutral interstellar medium. [CII] observations of DSFGs trace both the PDR and ionized gas \citep[e.g.][]{Spilker22}. However, their stellar distribution and ionized gas (without contamination of the PDR emission) properties remained hidden. Many of these galaxies are dark at HST wavelengths, or only reveal very faint emission. Only now with the arrival of JWST we are able to reveal the restframe optical and near-infrared wavelengths via imaging \citep{Colina23,Gillman23,Barrufet23,Gomez23,Cochrane23,Perez23CEERS,Zavala23,Akins23} and  spatially resolved spectroscopy, where we have for the first time access to emission lines originating in the ionized gas, tracing the most recent star formation \citep{Alvarez23,Arribas23,Jones23HFLS3,Parlanti23}. 

In this paper we analyze the ionized gas distribution and kinematics of the DSFG GN20. GN20 was identified as a bright 850 \micron\ source in GOODS-North based on deep SCUBA imaging \citep{Pope06}. GN20 is a DSFG at redshift 4.0548 \citep{Carilli11}, located in a proto-cluster or galaxy overdensity \citep{Daddi09} and has an infrared luminosity of 1.86\,$\times$\,10$^{13}$\,L$_{\odot}$. The star formation rate derived from fitting the observed spectral energy distribution (SED), assuming a Chabrier initial mass function (IMF), adds up to 1860 \msunperyear\ \citep{Tan14}. With a total stellar mass of 1.1\,$\times$\,10$^{11}$ \msun\ \citep{Tan14}, GN20 is located well above the $z$=4 star formation main sequence \citep{Speagle14,Caputi17}. The specific star formation rate  of GN20 computes to sSFR = $10^{-7.7}$ yr$^{-1}$, just below the starburst line for $z$ = 4 -- 5 derived by \citet{Caputi17}.

Deep Expanded Very Large Array (EVLA) observations of GN20 in the CO (2-1) line \citep{Carilli11,Hodge12} revealed a large, rotating molecular gas disk, visible under an inclination of 30\degr\
($M_\mathrm{dyn}$\,=\,5.4\,$\pm$\,2.4\,$\times$\,10$^{11}$\,M$_{\odot}$), extending as far at 14 kpc in diameter with a very clumpy spatial distribution. \citet{Hodge12} conclude that the presence of this disk would point to another process than a major merger as a driver for the starburst. This accretion could be enhanced by the interaction with the other galaxies in the overdensity. 

Recently \citet{Colina23} presented deep MIRI F560W imaging observations, probing the rest-frame near-infrared view of GN20. This study revealed for the first time the presence of a large stellar disk off-centred from the bright nucleus. This would, however, suggest that GN20 is involved in an interaction or a merger event, as supported by the presence of a secondary nucleus in the F560W image. 
 
In this paper we present  MIRI/MRS observations of GN20, aiming at detecting and characterizing the emission from the ionized gas, tracing the most recent star formation. The MRS observations are focused on the ionized hydrogen emission lines \paa\ ($\lambda_{\rm rest}$ = 1.8756 \micron) and \pab\ ($\lambda_{\rm rest}$ = 1.2822 \micron). Even though intrinsically fainter than H$\alpha$, these lines suffer significantly less from extinction and should therefore be easier to detect in galaxies with significant dust content such as DSFGs. In Sect. \ref{sec:obs} we present the observations, data reduction and post processing as well as the ancillary data. In Sect. \ref{sec:analysis} we outline our analysis methods. In Sect. \ref{sec:results} we present the line emission map and analyse the spectral and spatial distribution. Our discussion in the context of the evolution of DSFGs and our conclusions are presented in Section \ref{sec:discussion} and \ref{sec:conclusions} respectively.

 In this paper we adopt the cosmological parameters from the \citet{Planck15} and a redshift of GN20 of  $z$=4.0548. With these parameters, 1\arcsec\ is equal to 7.08 kpc. We use vacuum emission line wavelengths.

\section{Observations and data reduction}\label{sec:obs}

\subsection{MRS observations}

GN20 was observed by JWST on  November 24th, 2022 using the Mid-Infrared Instrument \citep[MIRI,][]{Rieke15,Wright15,Wright23} as part of the European Consortium MIRI Guaranteed Time (proposal ID 1264, PI: L. Colina). This paper presents the data obtained with the Medium Resolution Spectrograph \citep[MRS,][]{Wells15,Argyriou23}. \citet{Colina23} present the F560W imaging of GN20 obtained while MRS was performing background observations.

The MRS observations were obtained  in the MEDIUM configuration. The MEDIUM configuration provides observations in four different wavelength intervals; 5.66–6.63 \micron, 8.67–10.13 \micron, 13.34–15.57\micron, and 20.69–24.48 \micron\ \citep{Labiano21,Argyriou23}. This paper describes the analysis of MRS channels 1 and 2 in the MEDIUM configuration, our observations do not have enough sensitivity to detect any emission lines in channels 3 and 4, these bands are not discussed further.  
The observations mainly cover the \paa\ emission line ($\lambda_{\mathrm{rest}}$ = 1.87561 \micron), but channels 1 and 2 also cover additional ionized gas tracers as; \pab\ ($\lambda_{\mathrm{rest}}$ = 1.282 \micron), and [SiVI] ($\lambda_{\mathrm{rest}}$ = 1.963 \micron) as well as two H$_{2}$ lines tracing molecular gas, namely H$_{2}$ (1-0) S(7) ($\lambda_{\mathrm{rest}}$ = 1.747 \micron) and H$_{2}$ (1-0) S(3) ($\lambda_{\mathrm{rest}}$ = 1.957 \micron). 

The data were obtained in the SLOWR1 readout mode using a 4-point extended source dither pattern, repeated twice, resulting in eight different exposures. Each exposure contained one integration. Each integration contained 40 groups, resulting in  955.6 seconds per integration, summing up to a total of 7645 sec. Additionally, a background observation on an empty sky region, comprising of a 2-point dither with the same observing setup was taken to aid in the background subtraction.

\subsection{Calibration}

The MRS data were calibrated using the developers' version of the JWST calibration pipeline, version 1.11.1.dev3+g04332ea2 \citep{Bushouse23} with context 1097 of the Calibration Reference Data System (CRDS). We follow the same data reduction strategy as described in \citet{Alvarez23,Bosman23,Alvarez23JD1149}.

Stage one of the MRS pipeline \citep{Morrison23} is executed with mostly standard settings, producing the rate files. As our deep observations contain many groups and are taken in the SLOWR1 readout mode, special care has to be taken to remove the cosmic ray (CR) showers present in the data. The developers version we used contained the latest algorithms to minimise the effect of the CR showers. We turned on the \emph{find\_showers} keyword and, following \citet{Bosman23} set the \emph{time\_masked\_after\_shower} and \emph{max\_extended\_radius} to 60 seconds and 400 pixels respectively.  This procedure successfully removes a number of faint CR showers, however the brightest showers remain present.
 
From the rate files, we remove the warm pixels by calculating the median of all source and background frames. This median frame has all the cosmic ray events removed and only events persistent over many frames remain present. The source emission is very faint and not detected in each single frame.  To remove the warm pixels we applied sigma clipping and updated the data quality (DQ) frame of each individual rate file. The GN20 observations have a lower dark current than the reference dark present in the CRDS used by the pipeline. 
This results in an offset of the values of the rate files. We measure the median of the pixels in the inter-channel area of the detector and correct for the found offset ($\sim0.16$ DN/s). 

The second stage of the pipeline \citep{Argyriou23,Patapis23,Gasman23} is run with default parameters, apart from skipping the background subtraction step. 
After stage two we performed a pixel-by-pixel background subtraction. We calculate a master background frame by median averaging both object and background frames. Due to the different dither positions median average does not contain any source flux.

As a test we created a background frame from the two background observations and compared the rms in the datacube to that  created from the background with all the frames. The lowest rms was reached by using the median of all the frames as the background. We also extracted a spectrum of the \paa\ emission in both cubes and found the same, showing that no source emission is subtracted.  The background frame was subtracted from each object frame. Finally, stage three of the pipeline was run, skipping the \emph{master background subtraction} step. In the outlier rejection step we use a \emph{kernel\_size} of\, `11, 1' pixels, and  a \emph{threshold\_percent} of  99.8 \%. By combining all the exposures, 4 3D datacubes are produced (one for each band) using the \emph{drizzle} algorithm \citep{Law23}. This produces  datacubes with a spatial resolution of 0.13\arcsec per pixel (channel 1)  and 0.17\arcsec per pixel (channel 2). The FWHM of the MRS observations is calculated as 0.31$\times$\,$\lambda$ (\micron)/ 8\micron, resulting in 0.37\arcsec (2.6 kpc) at the wavelength of \paa\ \citep{Argyriou23}. The spectral resolution of the observation is R= 3774 for \pab\ and R=3390  for \paa\ \citep{Labiano21,Jones23}.

As already stated,  our data on GN20 are taken with the SLOWR1 readout and consist of long integrations, cosmic ray showers affect a significant part of the frames. Even though the updated pipeline significantly improves the removal of the CR showers, not all are removed and due to the faint emission of the target they considerably affect the extracted spectra to be analysed. The CR shower affect rather large parts of the detector and have elliptical shapes,  Due to their spatial extent on the rate files, cosmic rays manifest themselves as stripes in the $x$-direction in the reduced data cubes. As a final step of the data reduction process we employed the algorithm developed by \citet{Spilker23} to remove the striping due to the CR showers which are not removed by the pipeline.

\subsection{Archival datasets}

In this paper we compare the MRS observations with a multi-wavelength dataset to obtain a comprehensive picture of GN20. We use the CO(2-1) Very Large Array (VLA) observations presented in \citet{Carilli11} and \citet{Hodge12}, with a synthesized beam size of 0.21\arcsec, to compare the distribution and kinematics of the ionized gas to that of the molecular gas. We also use the 880 \micron\ continuum emission obtained with the Plateau de Bure Interferometer \citep[PdBI,][]{Hodge15}. These data have a synthesized beam size of 0.2\arcsec x 0.3\arcsec.  

To relate the ionized gas properties to the stellar distribution in GN20 we use the JWST/MIRI F560W imaging \citep[1.1 \micron\ rest-frame, ][]{Colina23}. Additionally we use HST imaging (PI: Faber, ID: 12442) with the ACS camera (F606W, F775W and F814W) and the WFC3 camera (F105W, F125W and F160W), tracing the UV-optical rest-frame emission. The reduced HST images were downloaded from the Mikulski Archive for Space Telescopes (MAST), astrometically aligned to GAIA DR3 \citep{GaiaDR3}. 

\subsection{Astrometry}

To improve the absolute astrometry of the observations we followed the recommended procedure and obtained simultaneous imaging in the F770W filter. This image was reduced using the JWST pipeline v1.8.2 and context \textit{jwst$\_$1030.pmap} following the procedure presented in \citet{Alvarez23}. The reduced  F770W image is astrometrically calibrated using GAIA DR3 \citep{GaiaDR3}, resulting in an astrometric accuracy of $\sim$ 70 mas. The offset from the initial astrometry provided by the telescope was calculated and applied to the header of the MRS data cube. 

Both the F814W image from HST and the F560W MIRI image are astrometrically calibrated with GAIA DR3  with an uncertainty less than $\sim$70 mas. The 880mm and CO(2-1) observations were phase referenced to quasars and their absolute astrometry is estimated to be a tenth of their synthesized beam, i.e. 20 mas. This makes the total accuracy of all the data (70 mas) well within the size of a pixel in the MRS data cubes. 

As a final check we stacked the cube of channel 2 in the wavelength direction to get a pseudo broadband image, detecting the bright nucleus seen in the F560W imaging \citet{Colina23} in the continuum. We found that the continuum emission in the MRS cube overlaps with the bright nucleus in the F560W image. 

\section{Analysis}\label{sec:analysis}

In this section we describe the procedure we have developed  to extract emission line spectra of both extended and point source emission. Additionally we describe the procedure to create the emission line maps we use to study the spatial distribution of the ionized gas.

\begin{figure}[!t]
  \includegraphics[width=\hsize]{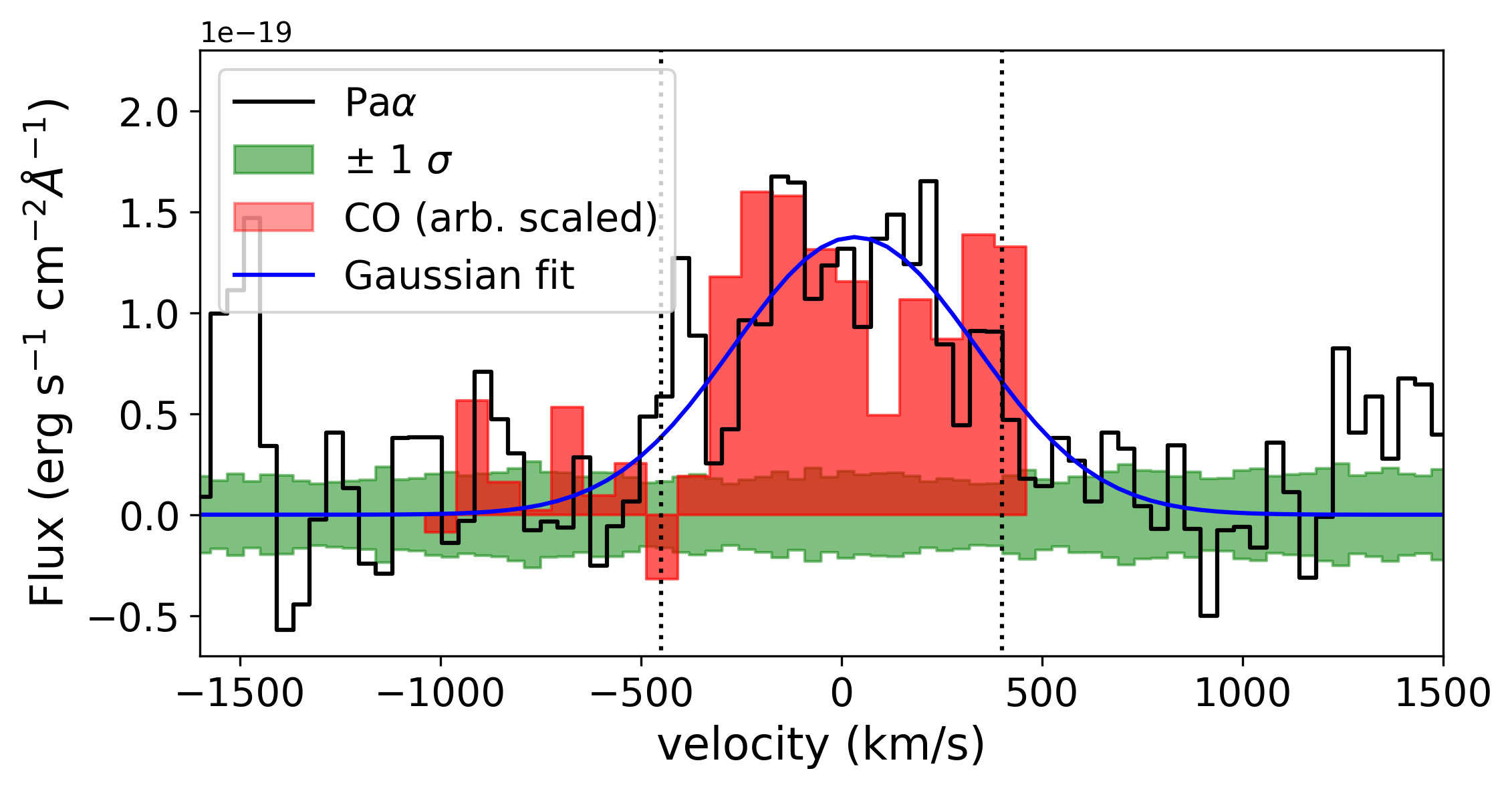}
      \caption{Integrated spectrum of GN20 centered on \paa. Plotted in green  is the 
      standard deviation derived from a background aperture with the same size as the galaxy spectrum. The spectrum plotted in red color shows the CO(2-1) emission extracted from the data presented in \citet{Hodge12}, overplotted in blue is the single Gaussian fit to derive the redshift (see text). The systemic velocity corresponds to a redshift of z = 4.0548 \citep{Carilli11}.}
    \label{fig:paa_spectrum}
\end{figure}

\subsection{Extraction of the spectra}

In order to get the integrated emission line profiles we extract the spectrum by integrating over a circular aperture using \emph{photutils} \citep{Bradley23} in each wavelength bin. For extraction of the integrated \paa\ emission line we use an aperture with a radius of 6 pixels (1.02\arcsec, 7 kpc), covering the observed extent of GN20 in CO(2-1) \citep{Hodge12} and the F560W continuum \citep{Colina23}. No aperture correction is applied as the aperture is relatively large compared to the size of the PSF (0.37\arcsec) and the emission is spatially extended. The resulting integrated \paa\ emission line profile is shown in  Fig. \ref{fig:paa_spectrum}.
We calculate the error by measuring the standard deviation (plotted in green) in a sky aperture with the same size, multiplied by the square-root of the number of pixels. We show in red the CO(2-1) spectrum of GN20  from \citet{Hodge12}. The two spectra look fairly similar, with the exception that the \paa\ emission shows blue shifted emission to $-500$ \kms, while the CO emission is only detected to $-400$ \kms.

To measure the redshift of GN20 from the \paa\ emission, we fit a single Gaussian profile to the integrated spectrum (Fig. \ref{fig:paa_spectrum}). We find $z_{\mathrm{Pa\alpha}}$ = 4.0553 $\pm$ 0.0006, within 1$\sigma$ from the value derived from the CO(2-1) observations by \citet{Carilli11}. For consistency with the literature we use $z$=4.0548 throughout the paper.

For the extraction of unresolved sources (clumps) we extract \paa\ spectra with a radius equal to the instrumental PSF (0.37\arcsec).  We measure the aperture correction for this aperture on the MRS over-sampled and drizzled PSF models from \citet{Patapis23}. We  find a correction factor of 1.42. For error determination of the point source extractions we extracted sky spectra at 9 places in the data cube where no galaxy emission is present and use standard deviation of these spectra as an error estimate.

\subsection{Construction of emission line maps}\label{sec:make_emissionmaps}

The integrated \paa\ spectrum (Fig. \ref{fig:paa_spectrum}) shows emission at velocities between $-450$ to +400 \kms. The CO(2-1) emission of GN20 shows a similar range of velocities and shows a velocity gradient due to the global kinematics of the galaxy \citep{Hodge12}. 

\begin{figure}[!t]
   \centering\includegraphics[width=0.65\hsize]{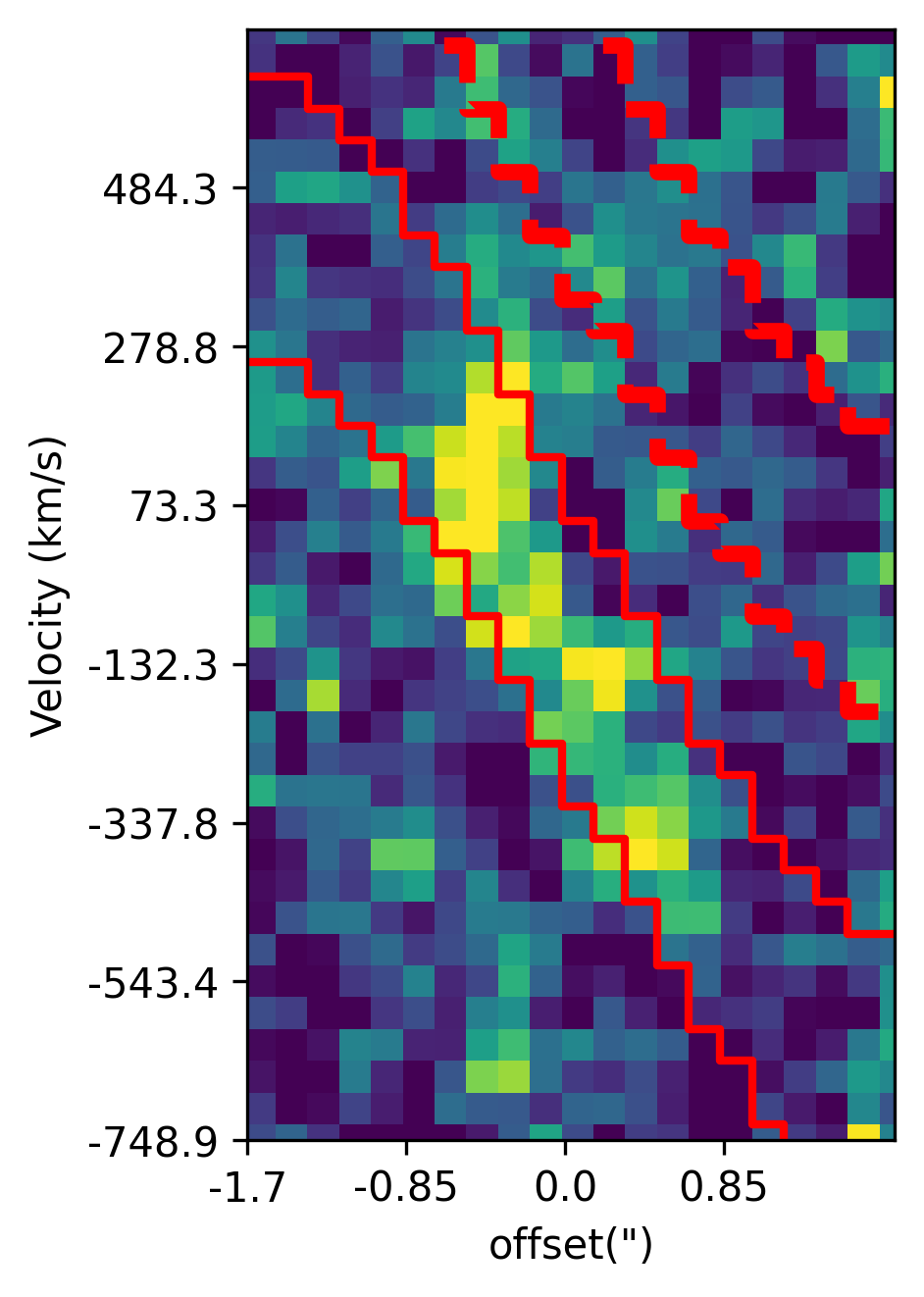}
      \caption{Position-velocity diagram of \paa\ along the major axis of the galaxy (PA = 25\degr). The offset is relative to the position of the bright nucleus of GN20 \citep{Colina23}. The red solid lines show the extraction apertures used for creating the \paa\ emission line map, while the red dashed lines shows the corresponding background map from which we derive the error on the emission line map.}
    \label{fig:paa_PV}
\end{figure}

In order to reduce the background noise in the emission line map, and keep the extraction window as narrow as possible, we first create a position velocity (PV) diagram using the \paa\ emission as guide. This line is the brightest and can be detected in the data cube. We sum up the emission over a pseudo-long slit placed over the major axis of the galaxy with a width of 8 pixels (1.4\arcsec). \citet{Hodge12} find a position angle (PA) of +25\degr. We find that the same PA gives the brightest \paa\ emission in the PV diagram (Fig.  \ref{fig:paa_PV}) and find no evidence for a difference in PA between the CO and \paa\ emission.

Based on the observed shape of the \paa\ emission in the PV diagram, we construct an aperture where the velocity scales as the arctan of the radius, commonly used to describe galaxy velocity profiles \citep{Glazebrook13}. We adjust the parameters such that all the \paa\ emission in the PV diagram is included. This function provides us with a central velocity of the \paa\ emission as a function of position along the major axis. From this we construct a map of central wavelengths of the \paa\ emission for each spatial pixel in the datacube. We construct the line map summing the flux in an aperture of 9 pixels in the wavelength direction (370 \kms) centered on the derived central wavelength. Parallel to the aperture centered on the \paa\ line we construct  a "background" aperture by shifting the same function with 10 pixels along the radius axis from which we derive the noise statistics.

As \paa\ is the brightest line, we use the \paa\ aperture shape  to extract line maps for  other expected emission lines. The H$_{2}$ and [SVI] lines are also located in channel 2. In order to extract emission line maps, we shift the map with the central wavelength derived for \paa\ to the central wavelength of the line of interest.  For the \pab\ aperture in channel 1, we scale the extraction aperture taking into account the different pixel size and spectral resolution between channel 1 and 2.


\section{Results}\label{sec:results}

\subsection{Emission line maps}

\begin{figure}[!t]
   \includegraphics[width=\hsize]{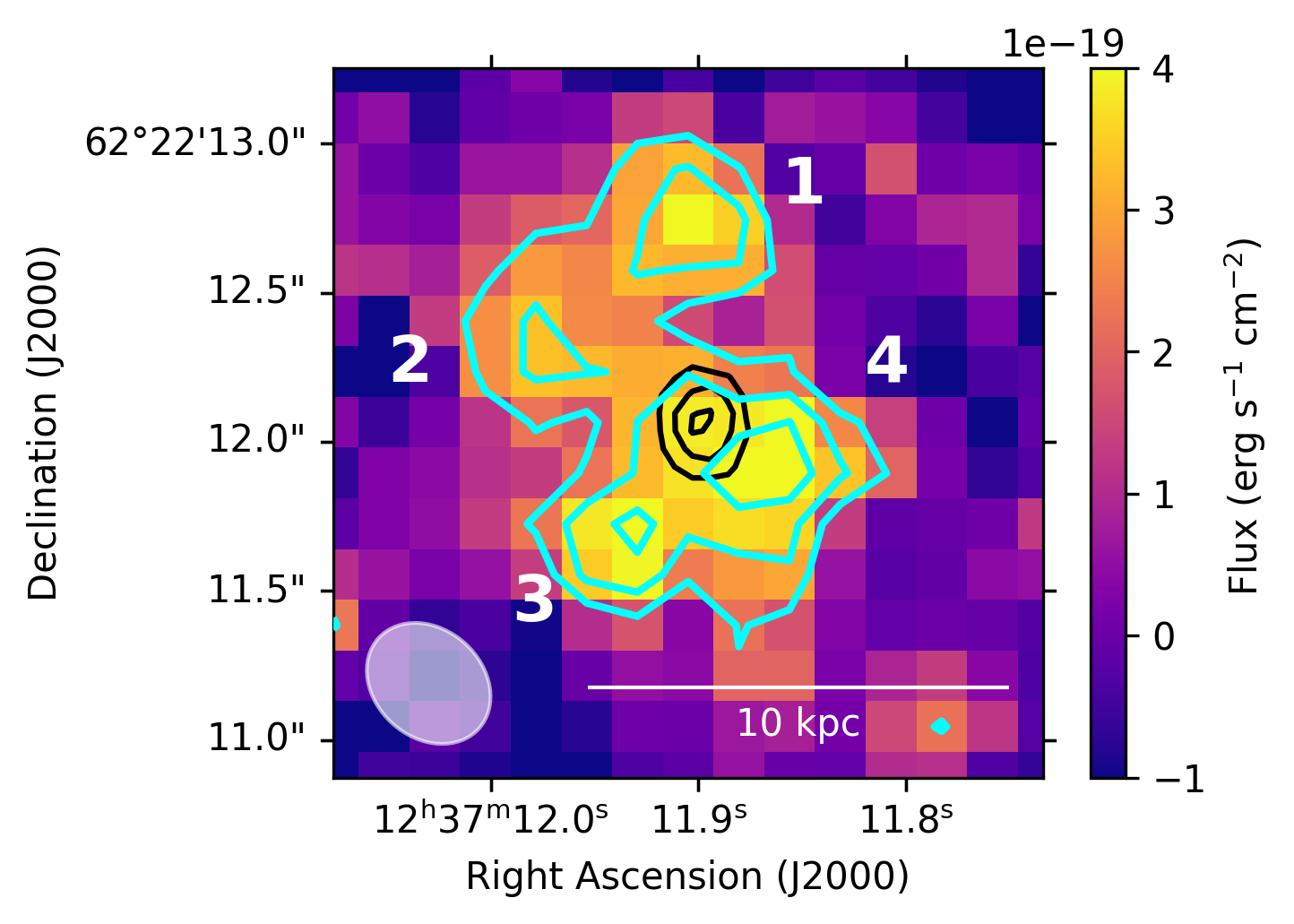}

      \caption{\paa\ emission line map after extraction based on the PV diagram (Fig.\ref{fig:paa_PV}, see text). The 2,3 and 4-$\sigma$ contours are shown in cyan. The black contours show the bright nucleus of GN20 detected in the F560W image \citep{Colina23}.
      The MRS point spread function at 9.47\micron\ is displayed as a grey circle.The PSF is slightly asymmetric and the size shows the FWHM in the $\alpha$ (x)- and $\beta$ (y)-direction, rotated according to the observed position angle.  The 4 clumps discussed in Sect. \ref{sec:clumps} are annotated.}
    \label{fig:paa_linemap}
\end{figure}

The extracted \paa\ emission line map is shown in figure \ref{fig:paa_linemap}.   We detect \paa\ emission in GN20 over a large extent of the galaxy, up to a radius of 6 kpc. The \paa\ emission shows a very clumpy structure with 4 clumps detected in the outer regions of the disk and very little emission towards the central regions of GN20. We do not detect any significant \pab\ emission or any of the other emission lines covered in the two channels.

We derive the error on the detected \paa\ flux by measuring the standard deviation in a large aperture in the "background" image as described in Sect. \ref{sec:make_emissionmaps}. For the \paa\ line map we find the standard deviation of the background image to be $1.06\times10^{-19}$ erg s$^{-1}$ cm$^{-2}$ per pixel, integrated over 9 wavelength elements. This would translate to a 3 $\sigma$ detection limit in an extraction aperture equal to the FWHM of the PSF (0.37\arcsec) of $1.2\times10^{-18}$ erg s$^{-1}$ cm$^{-2}$. Fig. \ref{fig:paa_linemap} shows the 2, 3 and 4-$\sigma$ contours of \paa, derived using this standard deviation. For \pab, we derive a standard deviation of $1.86 \times10^{-19}$ erg s$^{-1}$ cm$^{-2}$ per pixel, translating to a 3 $\sigma$ detection limit in a aperture equal to the FWHM (0.3\arcsec) of the PSF of $2.3 \times10^{-18}$ erg s$^{-1}$ cm$^{-2}$. For the other potential emission lines we derive the standard deviation in the same way and find 3 sigma detection limits of $1.5\times10^{-18}$ erg s$^{-1}$ cm$^{-2}$ for $[$SiVI$]$ and $1.4\times10^{-18}$ erg s$^{-1}$ cm$^{-2}$ for both the H$_{2}$ lines.


\subsection{Integrated star formation rate}\label{sec:sfrpaa}

We calculate the integrated \paa\ flux of GN20 by integrating the emission line map over a circular aperture with a radius of 6 pixels (1.02\arcsec). This results in a total flux of 1.9 $\pm$ 0.11 $\times$ 10$^{-17}$ erg s$^{-1}$ cm$^{-2}$ (Table \ref{tab:SFR_integrated}). We use the star formation rate calibration from \citet{Kennicutt12}. 
\citet{Kennicutt12} use a \citet{Kroupa01} IMF in their calibration, while in the remainder of the paper we use the IMF of \citet{Chabrier03}, yielding identical results \citep{Chomiuk11}.
We assume an intrinsic \paa\ over H$\alpha$ ratio of 0.116 \citep{Osterbrockbook} and we calculate a corresponding unobscured star formation rate (\sfrpaa) of 144 $\pm$ 9 M$_{\odot}$ yr$^{-1}$, assuming zero extinction.

\begin{table}[!ht]
\caption{Integrated flux measurements}
\centering
\begin{tabular}{lrr}
\hline\hline 
 Line  & Flux  & SFR$_{\mathrm{Av=0}}$ \\
	 & erg s$^{-1}$ cm$^{-2}$ & M$_{\odot}$ yr$^{-1}$ \\ 
\hline
\paa & $1.9 \pm 0.11 \times 10^{-17}$ & 144 $\pm$ 9 \\
\pab &  < $2.6\times 10^{-18}$ \tablefootmark{a}& < 43 \tablefootmark{a}\\
 \hline
\hline
\label{tab:SFR_integrated}
\end{tabular}
\tablefoot{\tablefoottext{a}{1 $\sigma$ upper limit}}
\end{table}

To derive an upperlimit of the \pab\ flux we integrate the flux in the \pab\ map in a similar circular aperture (8 pixels, 1.04\arcsec). This results in a 1-$\sigma$ upper limit for the integrated flux of \pab\ of 3 $\times$ 10$^{-18}$ erg s$^{-1}$ cm$^{-2}$ (equivalent to an \sfrpab\ < 43 M$_{\odot}$ yr$^{-1}$ in the case of no extinction). 

\begin{figure*}[!ht]
   \includegraphics[width=\hsize]{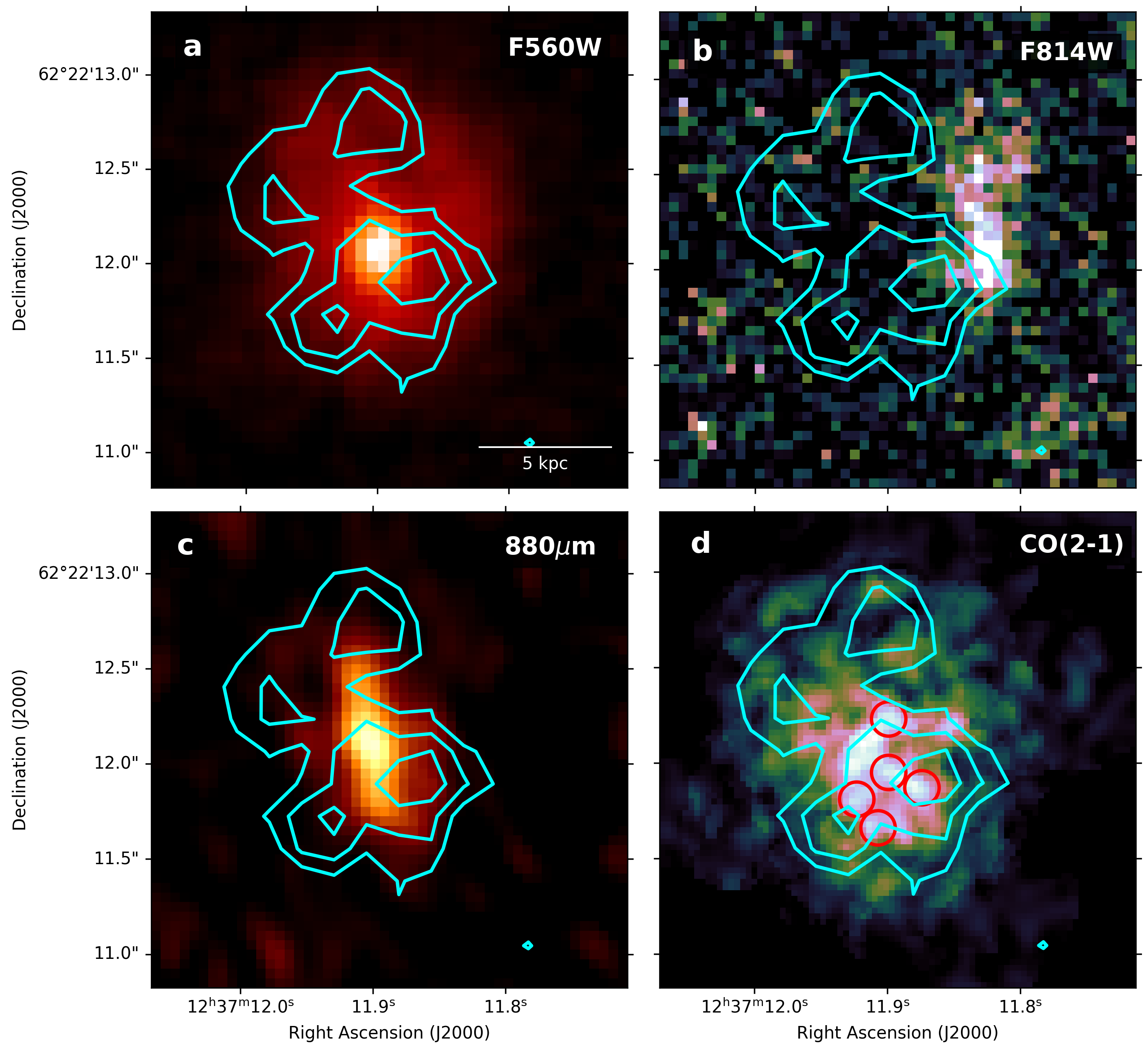}
      \caption{ Multiwavelength comparison of GN20 with the \paa\ emission line map (cyan contours). {\bf a}: F560W (rest-frame 1.1 \micron) image from \citet{Colina23}, {\bf b:} HST/ACS F814W (rest-frame 0.15\micron) image, {\bf c:} 880 \micron\ (rest-frame 160 \micron) image from \citet{Hodge15}, {\bf d:} CO(2-1) emission from \citet{Hodge12}. The red circles show the position of the 5 CO clumps identified by \citet{Hodge12}.}
    \label{fig:GN20_multiwave}
\end{figure*}

The 3 $\sigma$ detection limit of \pab\ is 7.89 $\times$ 10$^{-18}$  erg s$^{-1}$ cm$^{-2}$, comparing this 
to the observed \paa\ flux allows us to put a constraint on the foreground extinction. We assume an intrinsic \paa\ over \pab\ ratio of 2.075 \citep{Osterbrockbook}, for T = 10$^{4}$ K and an electron density (N$_{e}$) of 10$^2$ cm$^{-3}$ (for N$_{e}$ = 10$^4$, this value would be 2.051). The observed (3 $\sigma$ lower limit) \paa\ over \pab\ ratio is 2.37, a factor 1.08 higher than the theoretical ratio, this would result in a lower limit of  $A_V$= 4.3 mag, assuming a \citet{Cardelli89} extinction law with an R$_{V}$ = 4.05. Taking the 1 $\sigma$ error (Sect. \ref{sec:sfrpaa}), this lower limit in $A_V$ would increase to $A_V$ = 39.9 mag.  In Sect. \ref{sec:discussion_sfr} we compare the derived SFR to values derived via other tracers.


\subsection{Multi-wavelength comparison}

The \paa\ emission is observed over an area with a radius of $\sim$ 6 kpc (Fig. \ref{fig:paa_linemap}). The spatial distribution is very clumpy with most of the emission coming from the outer regions of the galaxy. In this section we compare the \paa\ spatial distribution with images tracing the stellar content, the cold dust, and the cold molecular gas in GN20. In Fig. \ref{fig:GN20_multiwave} we compare the morphology of the \paa\ line emission, represented by the cyan contours to the tracers discussed above.

 Comparison with the stellar distribution traced by the F560W image \citep[panel a,][]{Colina23} shows that the \paa\ emission has a similar spatial extent as the F560W emission. The \paa\ clumps we detect are all originating in the outer areas of the stellar disk. Towards the bright central point sources we detect very little \paa\ emission. North-west of the brightest clump 4 the 3 $\sigma$ contour shows an extension towards the nucleus (Fig. \ref{fig:paa_linemap}).

Panel b shows the comparison with the HST F814W image, tracing the rest-frame 0.15 \micron\ emission. We find that there is virtually no overlap between the \paa\ emission and the rest-frame UV emission. The F814W reveals the UV-bright part of the galaxy where extinction is expected to be very low. Surprisingly, the UV emission is not associated with \paa\ nebular emission (see Sect. \ref{sec:uvregion} for further discussion about the nature of the UV emission).

Comparison with the 880 \micron\ emission shows very little spatial correlation, most of the 880 \micron\ emission comes from the central area of the galaxy, while \paa\ emission is found further out.  Comparison with the CO (2-1) emission line map (panel d) shows again that the CO emission is mostly coming from the central area of the galaxy where \paa\ is not detected. The \paa\ emission overlaps with the fainter outer areas of the CO emission. Similar to the \paa\ spatial distribution, also the CO emission shows a clumpy structure, although the clumps are located further inwards \citep{Hodge15}.

As discussed in Sect. \ref{sec:discussion_sfr}, this multi-wavelength picture of GN20 suggests very high and strongly spatially varying extinction in GN20, where \paa\ and UV emission only emerge from the outskirts of the disk. Large extinction variations on (sub)kpc scales are commonly observed in local LIRGs and ULIRGs \citep[e.g.][]{Alonso06,Piqueras13}.

\begin{figure*}[!t]
   \includegraphics[width=\hsize]{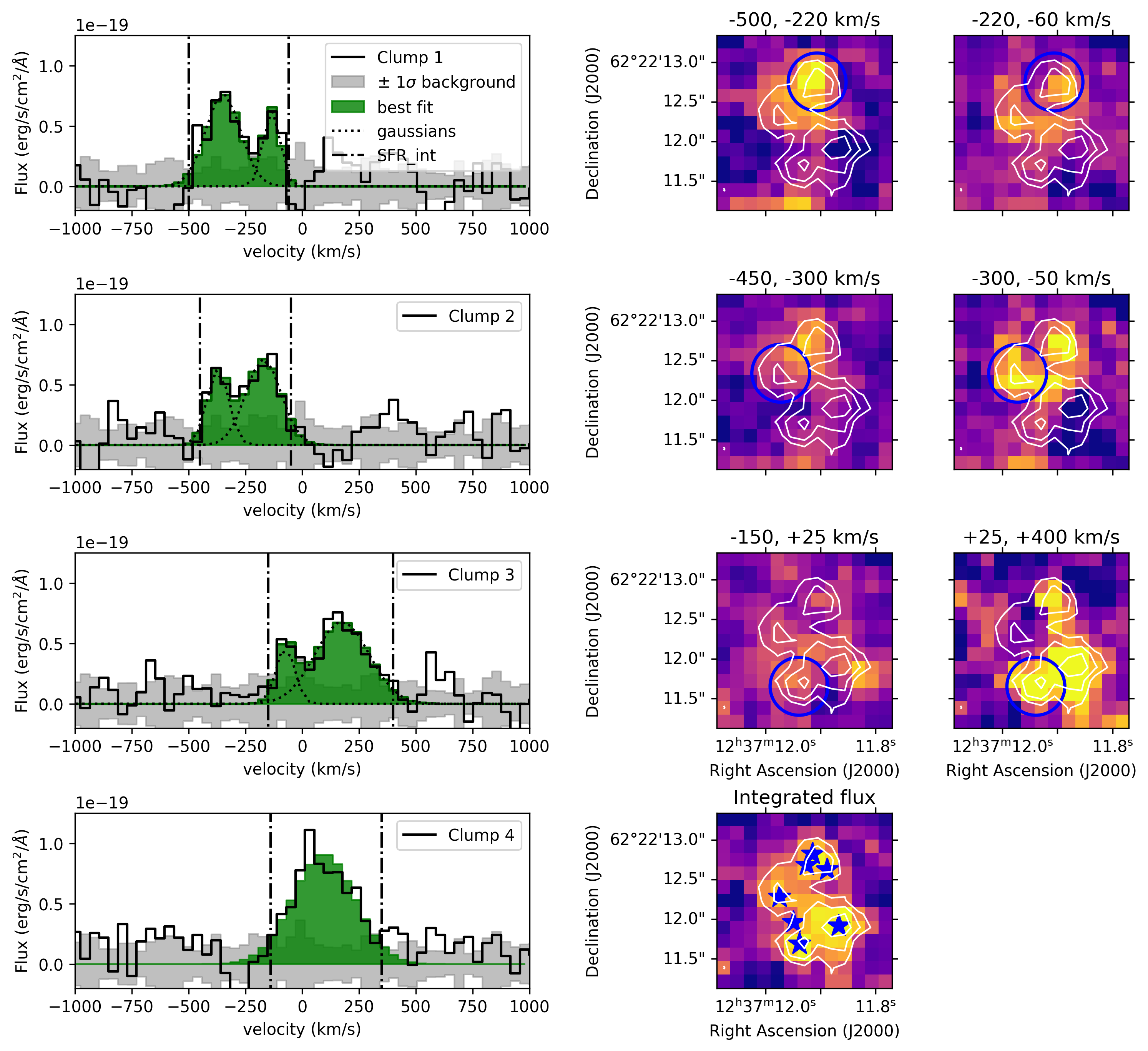}
      \caption{\emph{Left column:} Integrated spectra of the clumps identified in the \paa\ line map. The grey histograms represent the 1$\sigma$ errors. The vertical dotted lines represent the borders in which the flux and integrated star formation rate  is calculated (Tab. \ref{tab:clumps}). The green histograms show the best gaussian fits to the emission profile, with the dotted line the individual gaussians for clumps 1,2 and 3.
      \emph{Middle and right columns:} \paa\ channel map for each clump annotated by the blue circle. The two channel maps for each clump show the location of the two different peaks in the \paa\ spectra. Clump 4 is fitted by 1 Gaussian only and no channel maps are shown. In the emission channel map related to the red peak of clump 3 (row 3, right channel map), emission is visible at the location of clump 4 due to the similar velocity range. The peak at the location of clump 1 is caused by a (2-$\sigma$) peak at +125\kms in the spectrum of clump1.  The bottom image in the middle column displays the integrated \paa\ linemap (Fig. \ref{fig:paa_linemap}) with the location of the sub-clumps overlayed as blue asterisks.}
    \label{fig:paa_clumps}
\end{figure*}

\subsection{\paa\ clumps}\label{sec:clumps}

The \paa\ emission line map shows that the \paa\ emission is concentrated in four clumps, annotated in Fig. \ref{fig:paa_linemap}. We extract the spectra of the clumps with an circular aperture with a radius of  2.15 pixels (FWHM of the PSF at the observed wavelength of \paa).  The spectra are displayed in the left column of Fig. \ref{fig:paa_clumps}. All clumps show a spectrally resolved \paa\ emission profile, with clumps 1, 2 and 3 showing a double peaked profile. Clump 4 shows one single broad profile. 

We calculate the \sfrpaa\ of each of the individual clumps by integrating the spectrum over the velocity range where the clumps are detected (vertical dashed-dotted lines in Fig. \ref{fig:paa_clumps}). We find values between 48 and 70 \msunperyear\ (Table \ref{tab:clumps}). The total \sfrpaa\ by summing up the values of the clumps is 225 $\pm$ 42 \msunperyear, 55\% higher than the integrated \sfrpaa\ from the emission line map. 

The discrepancy between these two values could be caused by the fact that we applied an aperture correction to the clump spectra, but not to the integrated galaxy spectrum as that is not a point source and is extracted over a much larger aperture. Additionally, the clumps are located close to each other with an average separation of $\sim$0.5 -- 1.0\arcsec, making the wings of the PSF of the clumps overlap each other.
What we can conclude is that the clumps dominate the total emission in the galaxy and  no significant diffuse \paa\ emission is detected.


\begin{table*}[!ht]
\caption{Clump properties}
\centering
\begin{tabular}{lrrrrrrr}
\hline\hline 
 Clump  & R.A. & Dec & Flux  & SFR$_{\mathrm{Av=0}}$ & velocity & $\sigma$ & M$_{\mathrm{virial}}$\\
	 & &  &  ($10^{-18}$ erg/s/cm$^2$) & (M$_{\odot}$ yr$^{-1}$) & (\kms) & (\kms) & ($10^9$ M$_{\odot}$) \\
\hline
  1  & \textbf{12h37m11.91s} & \textbf{62\degr22\arcmin12.8\arcsec}& 6.6 $\pm$ 2.5 & 50 $\pm$ 19 & --- & --- & --- \\
  1a & 12h37m11.89s & 62\degr22\arcmin13.0\arcsec & 4.6 $\pm$ 3.1                        & 35 $\pm$ 24          & -348 $\pm$ 16 & 77 $\pm$ 16 &  < 6.8 $\pm$ 2.8 \\
  1b & 12h37m11.87s & 62\degr22\arcmin12.8\arcsec  & 1.9 $\pm$ 2.4                        & 14 $\pm$ 18         & -134 $\pm$ 16 & 37 $\pm$ 15 & < 0.05 $\pm$ 0.04  \\
  \hline
  2  & \textbf{12h37m11.98s}&\textbf{62\degr22\arcmin12.4\arcsec}  & 6.3 $\pm$ 2.4  &   48 $\pm$ 18 & & --- &  --- \\
  2a & 12h37m11.90s  & 62\degr22\arcmin12.8\arcsec & 2.3  $\pm$ 2.5                     & 18 $\pm$ 19         & -371 $\pm$ 20 & 47 $\pm$ 15 &  < 1.2 $\pm$ 0.8  \\
  2b & 12h37m11.87s & 62\degr22\arcmin12.9\arcsec & 4.4 $\pm$ 3.9                        & 33 $\pm$ 29        & -170 $\pm$ 21 & 77 $\pm$ 24 &  < 7.8 $\pm$ 4.8 \\
\hline
  3  &\textbf{12h37m11.94s} & \textbf{62\degr22\arcmin11.7\arcsec} & 7.5 $\pm$ 3.3  &    57 $\pm$ 25 & --- & --- & ---\\
  3a   & ---                         & ---    &  1.5  $\pm$ 2.0&   12 $\pm$ 15   & -74 $\pm$ 18 & 41 $\pm$ 15 &  < 0.4 $\pm$ 0.3  \\
  3b &12h37m11.91s &  62\degr22\arcmin11.9\arcsec&  6.4 $\pm$  3.7                       & 49 $\pm$ 28         & 175 $\pm$ 20 & 119 $\pm$ 23 &  < 19 $\pm$ 8  \\
\hline
  4 &\textbf{12h37m11.86s} & \textbf{62\degr22\arcmin11.9\arcsec} & 9.3 $\pm$ 3.1  &   70 $\pm$ 23 & 94 $\pm$ 18 & 138 $\pm$ 18 &  < 26 $\pm$ 7  \\
 \hline
\hline
\label{tab:clumps}
\end{tabular}
\end{table*}

To derive the kinematics of the clumps we fit Gaussian profiles to the extracted spectra using the \textit{astropy} modeling package. Clumps 1, 2 and 3 are fit with two gaussians simultaneously.  Clump 4 shows a hint of a second component on the red side of the \paa\ profile, but the components are too close to allow a double Gaussian fit. We calculate the velocity dispersion of the ionized gas by quadratically subtracting the width of the instrumental line spread function \citep{Labiano21,Jones23}. The FWHM of the instrumental broadening at 9.47 \micron\ is 88 \kms\ ($\sigma_{\mathrm{inst}}$ = 38 \kms). 
From the derived velocity dispersion (Table. \ref{tab:clumps}), we can derive the virial mass using the isotropic virial estimator \citep[e.g.][]{Hodge12}:

\begin{equation}
M_{virial}= \frac{C \sigma^{2}R_{g}}{G} (M_{\odot}).
\end{equation}

The observed velocity dispersion is represented by $\sigma$ in \kms,  the gravitational radius by $R_{g}$ in pc and the gravitational $G = 1/232$ pc (\kms)$^2$ $M_{\odot}^{-1}$ is the gravitational constant. The factor $C$ depends on the mass distribution and varies from $C = 7.4$ for an exponential profile to $C = 4$ for a Vaucouleurs profile \citep[e.g.][]{Bellocchi13}, introducing a factor two of uncertainty. In order to compare our results directly with the properties of the CO clumps, we follow \citet{Hodge12} and choose $C = 5$ (a uniform sphere). For the gravitational radius we use the FWHM of the MRS PSF at 9.48 micron (0.367\arcsec, 2.6 kpc). 
Given that the \paa\ clumps are not resolved, the effective radii derived in this way will only be upper limits, and the derived masses are thus also upper limits. The resulting values are presented in Table \ref{tab:clumps}. 

The velocity profile of the clumps suggests that they are not single massive star forming clumps, but they consist of at least two sub-clumps shifted in velocity. Even clump 4 shows a hint of a second component. To study the nature of the clumps in more detail we construct channel maps of the \paa\ emission for each of the three double peaked clumps (middle and right columns, Fig. \ref{fig:paa_clumps}).  We make channel maps for each of the clumps separating the emission of each of the components. In the top row of Fig. \ref{fig:paa_clumps}, we show first the extracted spectrum of clump 1 with the double peaked emission, the middle figure shows the channel map of the red peak between $-500$ and $-220$ \kms, while the right figure shows the channel map of the blue peak ($-220$ -- $-60$ \kms). In the figures we highlight the clump for which the channel maps are made by a blue circle. 
The same is shown for clump 2 and 3. In the case of clump 4 we cannot separate the two peaks and we do not show the channel map.

For all three clumps the emission in the red and blue channels are spatially slightly shifted from each other. 
We fit a two dimensional 2D gaussian profile in \textit{Qfitsview} to determine the center of the clumps.
In the case of clump 1 the blue-shifted emission (clump 1b, Table \ref{tab:clumps}) is shifted by 0.45\arcsec (3.2 kpc) to the north-east from the red-shifted emission (clump 1a). For clump 2, the redshifted emission (clump 2a) seems to be related to the redshifted emission in clump 1, while the blue shifted emission (clump 2b) peaks at the south-east of the extraction aperture. Clump 3a is very faint and we could not measure the position of this clump in the channel map. The bottom image in the middle column of Fig. \ref{fig:paa_clumps} shows the \paa\ integrated line map (Fig. \ref{fig:paa_linemap}) with superimposed the location of the individual sub-clumps. This analysis shows that the \paa\ clumps are not single, very massive clumps, but more complex giant star forming regions blended by the spatial resolution of the MRS spectrograph, using both spatial and velocity information we can start separating the clumps into their different sub-clumps.

We compare the properties of these clumps to the clumps found in CO(2-1). \citet{Hodge12} found five clumps in the CO emission, located in the central regions of GN20. The \paa\ clumps are located at further distances from the center (Fig. \ref{fig:GN20_multiwave}), apart from \paa\ clump 4, which spatially coincides with CO clump nr 3 \citep{Hodge12}.

Due to the difference in spatial resolution between the CO and \paa\ maps it is hard to associate more clumps to each other. ALMA and HST observations of the nearby LIRG IC 4687 shows a similar pattern, where a large fraction of the  clumps is detected in only one of the two tracers \citep[\paa\ or CO(2-1),][]{Pereira16}. Studying a larger sample of LIRGS, \citet{Sanchez22} reached similar conclusions.  

The kinematic properties of the \paa\ clumps are similar to those of the CO clumps. The derived velocities reflect the large scale rotation curve of GN20 (Sect. \ref{sec:kinematics}). The measured velocity dispersion range between $\sim$ 40 and 138 \kms. The largest value ($138 \pm 18$ \kms) is measured for clump 4. Inspecting the emission line profile suggests that clump 4 also has two components, but they are too close in velocity to each other to fit them with a double gaussian. This artificially broadens the single Gaussian fit. The velocity dispersion found for the CO clumps is $\sim$65 \kms, in the same range as our \paa\ clumps. The CO map has a slightly higher spatial resolution than our \paa\ map, but also the CO clumps are spatially unresolved, resulting in upper limits for the dynamical mass. Within the limitations of both data sets the properties of both sets of clumps are similar.


 \subsection{Ionized gas kinematics}\label{sec:kinematics}

\begin{figure}[!t]
   \includegraphics[width=\hsize]{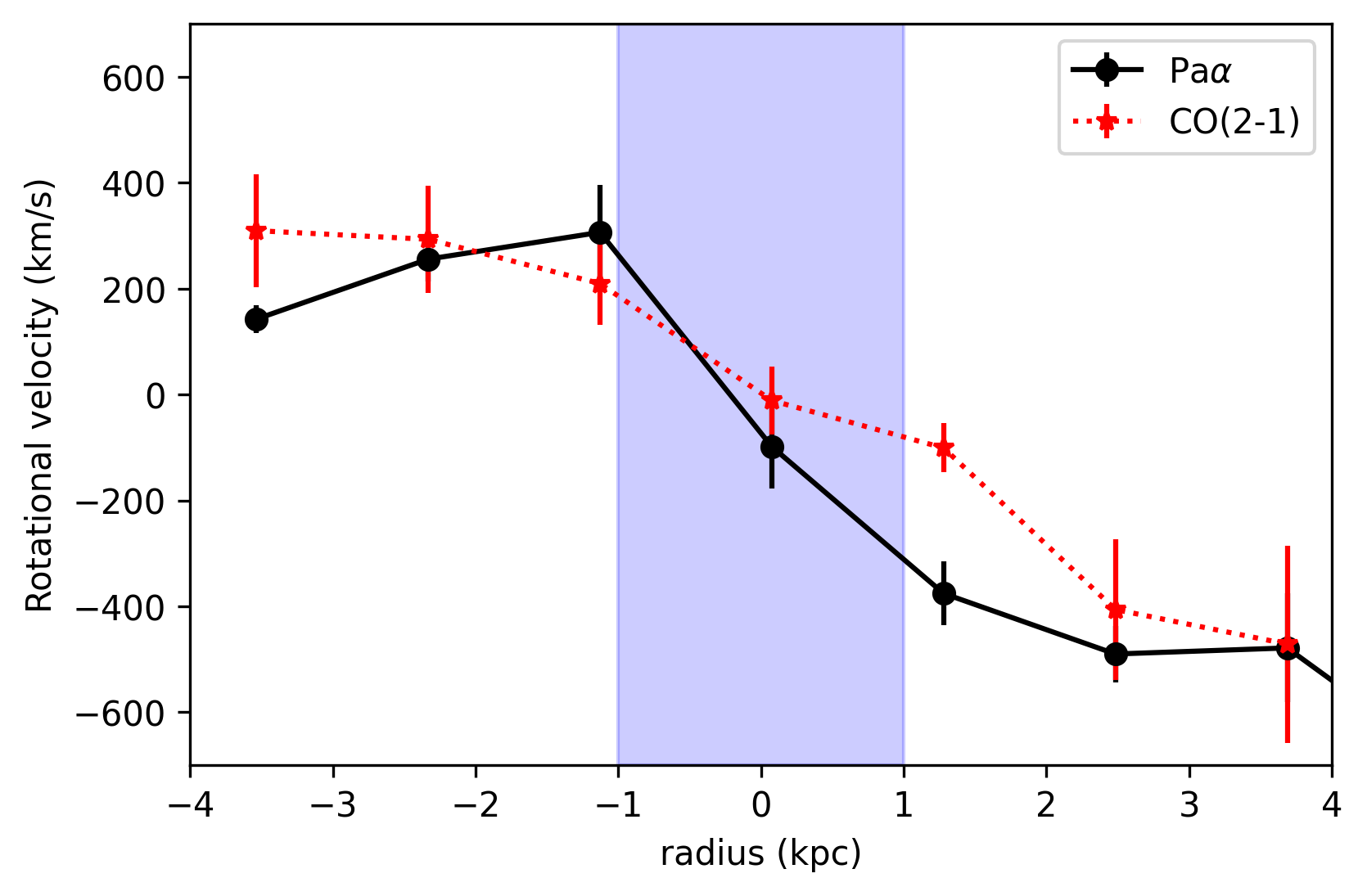}
      \caption{Rotational velocity of GN20 derived from a single gaussian fit to the \paa\ PV diagram (Fig. \ref{fig:paa_PV}) along the semi-major axis compared to the rotation profile from CO(2-1) derived from the velocity map in \citet{Hodge12}. The blue shaded area shows the range affected by beam smearing.}
    \label{fig:paa_vrot}
\end{figure}

We use the spatially resolved \paa\ emission to derive the global kinematics of the ionized gas in GN20 and compare that to the kinematics of the neutral gas as traced by CO(2-1). 
The signal-to-noise ratio of the \paa\ emission is too low to perform a Gaussian fit to each pixel in the data cube. We therefore use the PV-diagram (Fig. \ref{fig:paa_PV}), to derive the rotational velocity as a function of position along the semi-major axis of GN20. 

We construct the rotation curve along the semi-major axis with a position angle of 25\degr{} by fitting a Gaussian profile to each spatial pixel of the position velocity diagram (Fig. \ref{fig:paa_PV}). We correct the derived velocities for the inclination of GN20 of 30$\pm$15\degr (using the thin-disk approximation), as derived by kinematic modelling of the CO(2-1) emission \citep{Hodge12}, assuming the inclination of the CO is the same as \paa.  The center of the rotation curve (radius = 0) is determined by collapsing the PV-diagram along the wavelength direction. This results in the integrated flux along the semi-major axis of GN20 and reveals the bright nucleus as seen in the F560W image. We fit a Gaussian profile to the emission peak to determine the center of the rotation curve. 

The resulting rotation curve is shown in Fig. \ref{fig:paa_vrot}. We measure a velocity profile, increasing with radius to +300 \kms\ and \mbox{-400} \kms. The velocity dispersion varies between 85 and 190 \kms, with a typical error of 40 \kms.  The spatial resolution of the \paa\ observation is 2.6 kpc. This means that beam smearing strongly affects the shape of the observed velocity profile as well as the velocity dispersion. The gradual change of the rotational velocity in the central two - three kpc remains unresolved and cannot be distinguished from e.g. a step function.
The velocity dispersion is affected by overlapping clumps, as can be seen in the PV-diagram (Fig. \ref{fig:paa_PV}), where at $\sim$ \mbox{-0.7}\arcsec\ two clumps (clump 3 and 4) create an artificially broad \paa\ profile. 


As comparison we constructed the rotation velocity profile from the CO (2-1) data. We use the derived velocity map \citep[Fig. 3 in][]{Hodge12} to measure the rotational velocity. First we correct for the difference in spatial resolution between the CO(2-1) maps (0.19\arcsec) and the \paa\ map (0.365\arcsec). We convolve the CO(2-1) velocity map with a Gaussian kernel with $\sigma$ =  0.32\arcsec (the quadratic difference between the two FWHM values). Using the astrometric information,  we regrid the convolved velocity map to the same pixel scale as the \paa\ linemap to compare them on the same spatial resolution. We collapse the velocity map along the semi-major axis and use the maximum observed velocity at each distance as the radial velocity measurement, the standard deviation of the velocity values is used as estimate of the uncertainty. The resulting CO(2-1) velocity curve is plotted in red in Fig. \ref{fig:paa_vrot}.

Both the CO(2-1) and \paa\ rotation  curves reach the same minimum and maximum  velocity. However, they also show differences. The \paa\ rotation curve is much steeper than the CO(2-1) rotation curve, and both at -4 kpc as well as 1 kpc, the two rotation curves differ by $\sim$ 200 \kms. This could be related to the different spatial distribution of the \paa\ emission with respect to the CO(2-1) emission (Fig. \ref{fig:GN20_multiwave}.) The \paa\ clumps are situated around the nucleus and the \paa\ rotation curve is dominated by a rotating "ring" of clumps. The CO distribution is much smoother and stronger in the center and emitting more light at lower rotational velocities.

From the rotation curve of \paa\ we find a $v_{rot} = \frac{1}{2}(v_{max} - v_{min})$ = 550 $\pm$ 40 \kms.  This value agrees with the dynamical modeling of the CO(2-1) emission \citep[575 $\pm$ 100 \kms][]{Hodge12}. From the measured velocity dispersion we calculate the flux weighted velocity dispersion, a measure of the random gas motions in the galaxy \citep[e.g.][]{Glazebrook13}: $\sigma_{m}$ = 145 $\pm$ 53 \kms. This value should be interpreted as an upper limit to the real flux weighted velocity dispersion as we integrate the emission over the minor axis and overlapping clumps will create an artificial broadening of the emission line profile. 
From these quantities we calculate $v_{rot}/\sigma_{m} = 3.8 \pm 1.4$. This shows that the overall gas kinematics of GN20 are rotationally dominated in contrast to many (high-z) starburst galaxies \citep[e.g.][]{Forster09,Law09,Bik22}, which have $v_{rot}/\sigma_{m}$ values below the dividing line between dispersion and rotationally dominated of 1.83 \citep{Schreiber20}. \citet{Rizzo21} present a kinematic study of 5 strongly lensed DSFGs  around $z$=4.5 with ALMA and found high $v_{rot}/\sigma_{m}$ values for all galaxies, suggesting that they are rotationally dominated disk galaxies. Comparing GN20 values with those of (U)LIRGs shows that the  $v_{rot}/\sigma_{m}$ of GN20 is consistent with that of low-z LIRGs. \citet{Bellocchi13} derive an average  $v_{rot}/\sigma_{m}$ = 3.4 for LIRGs and $v_{rot}/\sigma_{m}$ = 2.0 for ULIRGs. \citet{Bellocchi13} divide their (U)LIRG sample in different morphological types corresponding to the different phases along the merging process. The measured $v_{rot}/\sigma_{m}$ makes GN20 consistent with  interacting systems with a perturbed disk. 

Considering that the \paa\ and CO(2-1) kinematics as well as the morphology suggest that GN20 is a large disk galaxy, we compare the derived value for $\sigma_{m}$ with theoretical predictions for $\sigma_{m}$ in the context of a galactic disk model. \citet{Krumholz18} shows that $\sigma_{m}$ in disk galaxies is determined by both star formation feedback processes and turbulence due to gravitational instabilities. 
Comparing the location of GN20 in the $\sigma_{m}$ vs SFR diagram \citep[Fig. 4 in][]{Krumholz18} show that the  high $\sigma_{m}$ values for GN20 derive from gravity driven turbulence and shares the location with ULIRGs in the more nearby universe \citep[see also][]{Arribas14}. Interestingly, these ULIRGs are not disk galaxies, but galaxies undergoing a major merger event. However, our derived $\sigma_{m}$ should be seen as an upper limit due to the overlapping clumps. Additionally, GN20 is significantly more massive and likely has a much more gas rich disk than ULIRG galaxies, resulting in higher $\sigma_{m}$ due to the higher mass of the clumps.

\section{Discussion}\label{sec:discussion}

Using deep MIRI/MRS spectroscopy, we have detected spatially resolved \paa\ emission in GN20. In this section we discuss the implications of our findings and compare them to auxiliary data. 

\subsection{Extinction in GN20}\label{sec:discussion_sfr}

We find a large discrepancy between the SFR derived from the \paa\ emission and the SFR derived from the infrared luminosity. Based on the \paa\ emission, under the assumption of no extinction, we derive  \sfrpaa\ = 144 $\pm$ 9 \msunperyear\ (Table \ref{tab:SFR_integrated}).  As shown in Sect. \ref{sec:sfrpaa}, the observed 3$\sigma$ upperlimit of \pab, does not provide a strict constraint on the extiction. \citet{Tan14} derived the infrared star formation rate (\sfrir) based on 1.2-3.3 mm continuum observations, assuming a \citet{Chabrier03} IMF,  and find a value for GN20 of \sfrir = 1860 $\pm$ 90 \msunperyear. This means that the observed \paa\ emission only reveals  7.7 $\pm$ 0.5 \% of the total star formation rate as detected from the mm observations. 

By comparing the \sfrpaa\ to the \sfrir\ for a sample of local LIRG galaxies, \citet{Tateuchi15} find that, after extinction correction, both SFR measurements agree within a scatter of 0.27 dex. \citet{Piqueras16} study the same relation, but with a sample of ULIRGs included. They find that in the case of the ULIRGs the \sfrpaa\ corrected for extinction is typically lower than the \sfrir. \citet{Gimenez-Arteaga22} find again a better correlation between the SFR derived from \pab\ with the \sfrir.

The disagreement between \sfrpaa\ and \sfrir\ can be caused by (a combination of) reasons: \emph{(i)} The mm emission is not (fully) caused by star formation, but has a contribution from an active galactic nucleus (AGN),
\emph{(ii)} Ionizing photons can be directly absorbed by dust, contributing to \sfrir\, but not to \sfrpaa\ \citep{Piqueras16},
\emph{(iii)} \sfrir\ traces the star formation history over a longer time (up to 100 Myr) than \paa\ (10 Myr) \citep{Kennicutt12}, and \emph{(iv)} The extinctions are so large that even \paa\ does not trace the full amount of ionized gas \citep{Alvarez23}.

A significant fraction of  SMGs are major mergers \citep[e.g.][]{Gillman23}, as their local ULIRG counterparts, which makes the presence of an AGN in the center a realistic possibility. GN20 is located in an overdensity and shows evidence for a double nucleus, suggesting that GN20 is interacting or in an advanced stage of a major merger \citep{Colina23}. Therefore the presence of an AGN can be expected \citep{Ricci17,Blecha18}. Based on the detection of the PAH 6.2 \micron\ emission, \citet{Riechers14} conclude that the infrared emission from the nucleus could contain a significant contribution of a buried AGN. The total observed infrared SED is, however, dominated by the obscured starburst. Also \citet{Colina23} attribute the bright  nuclear emission seen in the F560W imaging to a strong nuclear starburst. 

The center of GN20 contains a large amount of dust and gas, therefore a fraction of the ionizing photons will be absorbed by dust directly, before ionizing the gas.  
Additionally, \sfrpaa\ traces the most recent star formation over the last 10 Myr, while \sfrir\ is sensitive to star formation over a much longer time period as  intermediate mass stars  are also capable of heating the dust, but do not produce ionizing photons in sufficiently large quantities. \citet{Piqueras16} attribute the observed difference between \sfrpaa\ and \sfrir\ to both these effects. The difference between \sfrpaa\ and \sfrir\ in GN20 is, however, much larger than observed by \citet{Piqueras16}. \citet{Hodge15} shows that the obscured, strong, starburst in the center of GN20 has a gas depletion time of $\sim$ 100 Myr. If no \paa\ would be emitted from such a starburst, that would mean that the extreme starburst in the center suddenly stopped forming stars $>$10 Myr ago, while there is still a huge reservoir of molecular material present. 

Therefore, we attribute the discrepancy to the fact that we do not trace all the star formation with \paa\ due to the high extinction. As we do not detect any other hydrogen recombination line, we cannot derive the extinction for \paa. We can, however, derive an estimate from the comparison between \sfrpaa\ and \sfrir. Using the \citet{Cardelli89} extinction law with a R$_{V}$ = 4.05, we find a total extinction of A$_{V}$ = 17.2 $\pm$ 0.4 mag. As a more realistic approach we use an extinction model assuming the gas and stars are mixed. We use the relation derived by \citet{Calabro18} for a sample of dusty starburst galaxies at $z$=0.5-0.9 in combination with the extinction coefficients from \citet{Cardelli89} extinction law to get an estimate of the attenuation. Using equation 1 in \citet{Calabro18}, our ratio of observed SFRs gives a total visual attenuation of A$_{V,mixed}$ = 44 $\pm$ 3 magnitudes.  

The multi-wavelength comparison shown in Fig. \ref{fig:GN20_multiwave} shows that the mm continuum emission is concentrated in the center of GN20, while \paa\ is originating from the clumps at a few kpc distance from the center. Without extinction correction these clumps account for 7.7\% of the total star formation. An extinction corrected \sfrpaa\ of the clumps will increase this fraction, for example increasing this fraction to 20\% would require an $A_{V}$ = 7 mag for the \paa\ clumps. 

We attribute the remaining fraction of the star formation to be in the center, which would suggest an even higher extinction in the center of GN20. \citet{Piqueras13} show that most (U)LIRGs display a similar behaviour where large extinction variations, with areas with A$_{V}$ as high as 20-30 magnitudes, are found \citep[see also][]{Bohn23}. These (U)LIRGs typically have an obscured center.  However, in most local (U)LIRGs the extinction corrected SFR traced by the Paschen lines follows pretty well the \sfrir \citep{Gimenez-Arteaga22}, suggesting that  extinctions in most (U)LIRGs are not as extreme as in GN20.  Two ULIRGs  show very high extinction in their nuclei. Based on near-infrared extinction measurements, \citet{Engel11} derive very high extinction in the two nuclei of Arp 220. They show that some of the gas is so obscured that it is not detected in the near-infrared emission lines. Based on MIR diagnostics, \citet{Haas01} derive values of A$_{V,mixed}$ = 110 mag for the central starburst in Arp 220 and A$_{V,mixed}$ = 50 mag in the center of UGC5105 \citep[see also][]{Sturm96}. 
This shows that such high extinction as seen in GN20 is also detected in the more extreme ULIRG galaxies, where a very dusty extreme starburst is driven by the merger process. 

 The high extinction measured towards GN20 will have strong implication on stellar mass determinations from optical and near-infrared rest frame images. \citet{Colina23} derive a disk mass of $5.2\times 10^{10}$ \msun\ (assuming zero extinction), a factor of 2 lower than the stellar mass derived from SED fitting \citep{Tan14}. To quantify the effect of the spatially variable extinction on the stellar mass determination of GN20 is beyond the scope of this paper. The extinction most likely affects the youngest populations the strongest, while the older population is carrying the mass of the galaxy. A detailed multi-wavelength analysis utilizing HST, future NIRCAM and MIRI imaging covering the UV to near-IR rest-frame will address this issue. 

As shown in Fig. \ref{fig:GN20_multiwave} the different tracers in GN20 show  a very different spatial morphology. The high extinction makes the galaxy almost invisible and results in a significant suppression of the \paa\ emission from the central starburst. This would bias the measured half-light radii to larger values, affecting the mass-size relation of starburst galaxies \citep[e.g.][]{Ward23,Costantin23,Ormerod23}. Observations of local LIRGs show that the galaxies are much more compact in CO than in ionized gas or stars \citep{Bellocchi22}.
Additionally, \citet{Bellocchi22} compare the effective radii from \ha\ and \paa\ and find that the \ha\ radii are significantly larger, suggesting that \ha\ is more affected by extinction in the center of the LIRGS. 

 \citet{Fujimoto17} find similar trends based on HST and ALMA observations of $z$ = 1 -- 5 starburst galaxies. \citet{Popping22} demonstrate that  galaxies in the TNG50 cosmological simulation have larger half light radii in the optical and near-infrared light compared to the dust emission and attribute this difference to extinction.

High spatial resolution observations of U(LIRGs) have revealed extremely compact (<100 pc) obscured nuclei (CONs) in many ULIRGs and LIRGs \citep[e.g.][]{Aalto15,Donnan23}. Deeply buried in these CONs is either a rapidly growing super massive black hole  or a compact nuclear starburst, hidden behind very high column densities of dust and gas. In GN20 we do not have the spatial resolution to find such a compact region, however, the highly obscured and extreme nuclear starburst we find in GN20 might resemble what we observe as CONs in nearby (U)LIRGs.  This is consistent with the findings of \citet{Cortzen20} who show that the emission longward of 170 \micron\ is optically thick, consistent with a very compact starburst in the nucleus of GN20. Similar properties are observed for other high-redshift starbursts \citep{Jin22}.

\begin{figure}[!t]
   \includegraphics[width=\hsize]{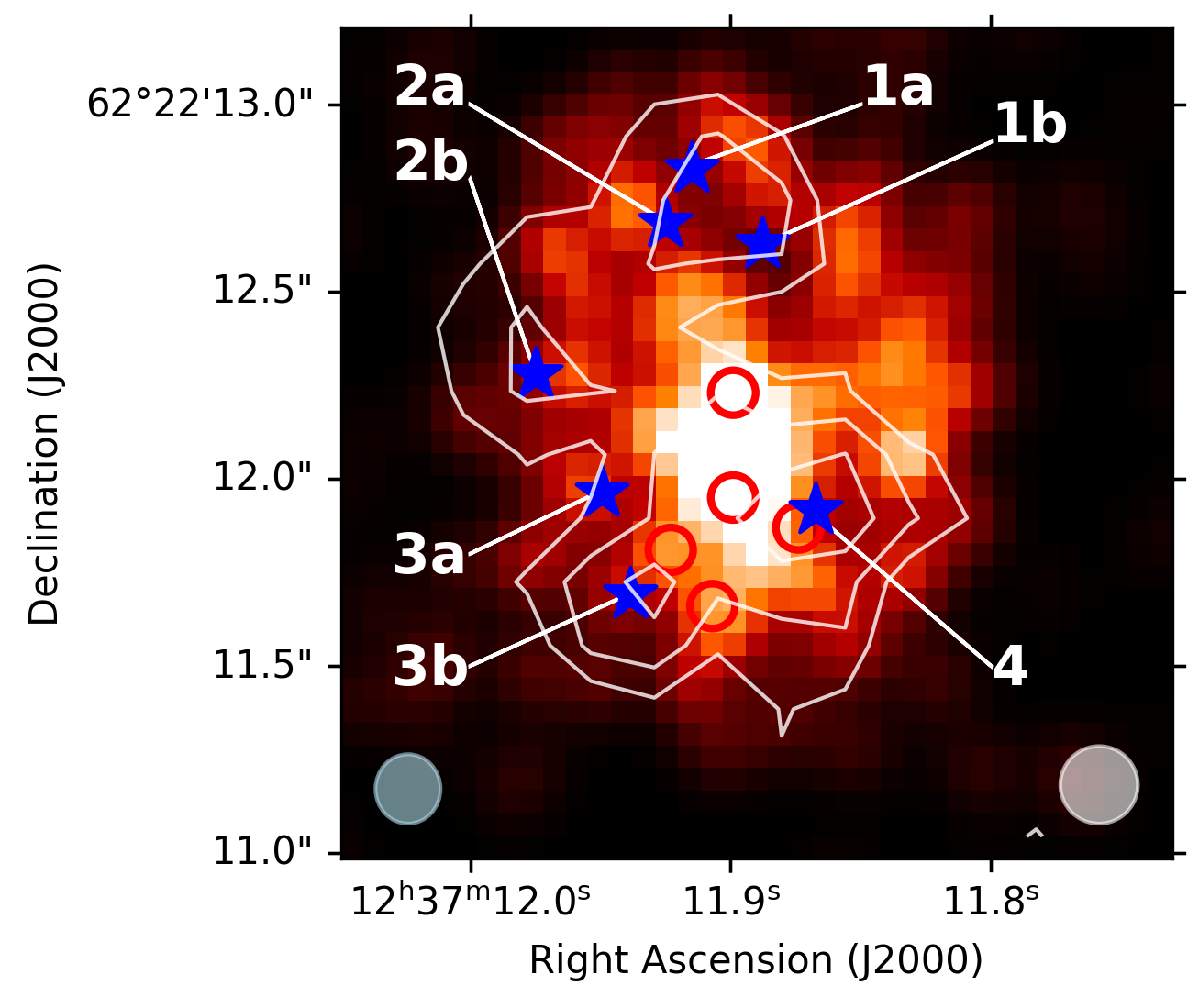}
      \caption{Convolved image of GN20 in F560W. The white contours stand for the contours of the \paa\ emission line map (Fig. \ref{fig:paa_linemap}). The blue asterisks show the position of the clumps identified in Sect. \ref{sec:clumps}. The red circles show the location of the CO clumps identified by \citet{Hodge12}.  The beam size of the CO observations is shown in the bottom left and the PSF of the MIRI F560W image on the bottom right.}
    \label{fig:F560W_deconv}
\end{figure}

\subsection{Nature of the star forming clumps}

We identify four individual clumps based on the \paa\ emission line map, using their kinematics we find that they are not single clumps, but consist of at least two sub-clumps in each of them. We search for continuum emission from these clumps in the F560W image \citep{Colina23}. The F560W image has a spatial resolution of 0.24\arcsec (1.7 kpc), slightly higher than that of the \paa\ linemap. 

Following \citet{Colina23}, we apply the Lucy-Richardson deconvolution algorithm \citep{Lucy74} to the image data. For the deconvolution of the input data, we use the empirical PSF derived by \citet{Colina23} which is based on close-by stars in the MIRI imaging field of view. Due to the confusion-free detection of GN20 in the MIRI data, we do not apply any background subtraction. In total, we use 10000 iteration steps to minimize the chance of artifacts. In the final step, we convolve the resulting delta map with a 3-pixel Gaussian Kernel filter adapting the procedure described in \citet{Peissker22}.
Figure \ref{fig:F560W_deconv} shows the resulting image with the \paa\ emission map contours in white. The blue asterisks show the different \paa\ clumps identified in Sect. \ref{sec:clumps} and the red circles show the position of the CO clumps identified in \citet{Hodge12}.
The final convolved F560W image shows a very clumpy structure, similar to the residual image \citep[Fig. 1 in][]{Colina23}, after removing the best fitted two component model, representing the smooth light distribution in GN20. Considering the clumpy nature of both the CO and the \paa, both tracing current star formation, the diffuse component of the F560W image could be related to more mature stellar populations having a smoother spatial distribution.

We find that some of the \paa\ clumps can be attributed to clumpy emission in the F560W image, clumps 2b, 3a and 3b show a matching F560W clump. Based on their location and kinematics we suggest in Sect. \ref{sec:clumps} that clump 1a and 2a might be tracing the same clump. In the F560W image, there is a clumpy structure located north-east of clump 2a, making this a potential counterpart. For clumps 1a and 4 on the other hand, no clear counterpart in the F560W can be found. Future higher spatial resolution NIRCAM imaging of GN20 will reveal more details on the clumpy structure of the GN20 disk.

Clumps are commonly observed in star forming galaxies at z $\geq$ 1 \citep[e.g.][]{Elmegreen09,ForsterSchreiber11,Livermore12,Adamo13,Dessauges18}. The clumpiness of high-$z$ galaxies is attributed to the higher gas fraction in  galaxy disks  \citep[e.g][]{Genzel08,Girard18}. The disks are fed by infall and accretion of gas-rich material, where clumps form via violent disk instabilities \citep{Dekel09,Ceverino10,Mandelker14,Inoue15}. In this scenario, the giant clumps that form have masses of a few percent of the mass of the disk \citep{Ceverino10}. \citet{Hodge12} derive a dynamical mass for GN20 of 5.4 $\pm$ 2.4 $\times$ 10$^{11}$ \msun. The upper limits we derive on the clump masses ($\sim$ 10$^{9}$ \msun, Table \ref{tab:clumps}) are consistent with the predictions of the violent disk instabilities scenario. The masses derived for the CO clumps by \citet{Hodge12} are consistent with this as well. Similarly, \citet{Spilker22} measure high velocity dispersion in the clumps detected in [CII]  in the merging DSFG SPT0311-58. This  also suggests very massive clumps under the assumption that the measured velocity dispersion is driven by gravitational motions. Higher spatial resolution observations might well break the observed clumps down in several sub-clumps \citep[e.g.][]{Messa22, Mestric22}, making the measured velocity dispersion only upper limits.

Additionally, merging galaxies trigger strong starburst resulting in the formation of clumps.  (U)LIRGs in the local universe show  clumpy star formation. \citet{Piqueras16} finds clumps in \paa\ in local (U)LIRGs of sizes between 300 pc and 1500 pc in the ULIRGs and a bit smaller in the LIRGs, while \citet{Larson20} find \paa\ clumps in local LIRGs and find sizes between 90 and 900 pc.

\subsection{Nature of the UV-bright region}\label{sec:uvregion}

The HST images of GN20 (Fig. \ref{fig:GN20_multiwave}) reveal a UV-bright elongated structure west of the center of the galaxy. It is detected in HST images, from F606W (rest-frame 0.12\micron) to F160W (rest-frame 0.32\micron).
 Also in the MIRI F560W image the arm is visible \citep{Colina23}. The convolved F560W image reveals 4-5 clumps associated with the UV-bright emission. This arm is not detected in CO, 880\micron\ and \paa. The fact that this region is detected in the UV would suggest low extinction. However, UV-bright would also imply a young stellar population, and therefore we would expect \paa\ from this low extinction, young region in GN20.  

Based on fitting of the observed UV to the radio spectral energy distribution (SED) of GN20, \citet{Tan14} find an \sfrir/\sfruv\ of $\sim$13, this would lead to an \sfruv\ of 140 \msunperyear. 
However, the UV emission only traces a small piece of the galaxy, therefore we reanalyze the HST data, by measuring the integrated flux with an elliptical aperture of 1 arcsec$^2$ in size, matching the size and morphology of the UV-bright emission (1.6\arcsec$\times$0.8\arcsec, PA: 0\degr).

Under the assumption of no extinction we calculate the \sfruv\ from the measured F814W flux using the relation detailed by \citet{Murphy11} and find \sfruv = 15 $\pm$ 2 \msunperyear. As a next step we perform a fit using CIGALE \citep{Burgarella05,Noll09,Boquien19} to the integrated UV-optical SED as seen by HST. We use a \citet{Chabrier03} IMF with a \citet{Calzetti00} extinction law and model the SED with two stellar populations; a young population ($<$ 40 Myr) with constant star formation history and a older single stellar population between 100 - 500 Myr. For the young stellar population (relevant for the UV emission) we find a SFR = 135 $\pm$ 70 \msunperyear\ and an extinction of A$_{\mathrm{V}}$ = 0.8 magnitudes. 

Next, we measure the flux of the UV-bright region in the MIRI F560W \citep{Colina23} and F770W (Crespo G\'omez et al, in prep) with the same aperture as for the HST images. We measure the flux in the images with the 2 component \textit{Lenstronomy} \citep{Birrer18} fit removed \citep[see][for details]{Colina23}. Repeating the CIGALE fit with the two extra MIRI points gives a similar SFR = 160 $\pm$ 100 \msunperyear\ for the young population, with a similar extinction as the previous fit. 

Based on the standard deviation derived for the \paa\ image (Sect. \ref{sec:sfrpaa}) we derive a 3 $\sigma$\ detection limit of 2.5 \msunperyear per pixel. Assuming a constant spatial distribution and an MRS pixel size of 0.029 arcsec$^2$ gives a 3$\sigma$\ upper limit of 15 \msunperyear\ for a 1  arcsec$^2$ aperture. This \paa\ upper limit is consistent with the directly measured \sfruv, but only under the assumption of no extinction. The CIGALE fits show a significant extinction of A$_{\mathrm{V}}$ = 0.8 magnitudes, increasing the \sfruv\ significantly. Under the assumption of the A$_{\mathrm{V}}$ = 0.8, the UV derived star formation rate would increase to 43 \msunperyear.

In summary, \paa\ traces the star formation over the last 10 Myr in GN20, while the star formation traced by the UV emission is sensitive to the last 100 Myr \citep{Murphy11,Kennicutt12}. 
The derived SFR values are consistent with a stellar population older than 10 Myr and younger than 100 Myr, such that it results in a bright UV region without strong ionized gas emission.

This is comparable with what is observed in the local LIRG galaxy NGC 1640 \citep{Adamo20}, where very little H$\alpha$ emission is detected towards a UV-bright arm. \citet{Adamo20} find a cluster population in this arm of $\sim$ 20-30 Myr. This population still contributes strongly to the UV flux, but is too old for significant production of ionizing photons.

\subsection{What triggered the starburst in GN20?}

 Two main scenarios have been proposed in the literature for the presence of starbursts in DSFG. The first scenario is that of a major merger, similar to local ULIRGS, while the second scenario is  violent disk instabilities in gas rich galaxy disks. In this section we summarize the properties of GN20 derived in this and previous papers and speculate about the origin of the starburst in GN20.

The offset between the nucleus and the center of the outer isophotes of the disk as well as the secondary nucleus found in GN20 by \citet{Colina23} would suggest that GN20 is at the final stage of a major merger. The extremely obscured nucleus  (Sect. \ref{sec:discussion_sfr}) and the very efficient star formation \citep{Hodge15} make GN20 display similar properties to the most extreme ULIRG galaxies. In this scenario, the star forming clumps would then be a byproduct of the merger process as seen in local merging (U)LIRGs \citep[e.g.][]{Pereira16,Piqueras16,Adamo20}. Additionally, GN20 is known to be located in an overdensity \citep{Daddi09}, making it likely that the star formation is at least affected by  gravitational interaction.

The kinematics of both the CO and \paa\ are consistent with a rotationally supported, gas-rich disk. 
Additionally, our \paa\ observations  and the CO(2-1) observations of  \citet{Hodge12} show that the star formation happens over a much larger spatial extent than in ULIRG galaxies \citep{Hodge16,Bellocchi22}, and covers the entire observed disk of GN20. These observations would hint that GN20 is a disk galaxy whose star formation is fuelled by accretion of neutral gas from intergalactic matter, where the clumps would form in the gas rich disk by violent disk instabilities \citep{Dekel09,Ceverino10,Inoue15}.

\citet{Hung15} study how the kinematics of a sample of (U)LIRGS would look like when redshifted to z=1.5 with a spatial resolution of 900 pc. They show that the kinematics of galaxies in a late stage of the merger process are indistinguishable from a disk-like rotational profile. 
On the other hand \citet{Ueda14} studied 24 local (U)LIRGS and find that extended molecular disks are common in the late stages of mergers. This is consistent with major merger simulations of \citet{Springel05}, predicting the formation of extended gas rich disks at the end of the merging process.

Beam smearing, due to the spatial resolution of the MRS, prevents us from deriving a well sampled rotation curve from the \paa\ kinematics, and makes it impossible to determine whether the kinematics are determined by a real gas rich galaxy disk, possibly rebuilt after the major merger, or a late-stage merger mimicking a disk. In the case of a rebuilt gas disk during the merger processes \citep{Ueda14}, the \paa\ clumps could possibly be formed in the disk due to violent disk instabilities, while the central starburst is the result of the actual merger.







\section{Conclusions}\label{sec:conclusions}

We present deep MIRI/MRS spectroscopy of \paa\ and \pab\ of the dusty star forming galaxy GN20 at z=4.0548. We detect for the first time ionized gas \paa\ emission out to a radius of 6 kpc, distributed in a clumpy morphology. The \pab\ line is not detected. From the integrated \paa\ flux we derive a \sfrpaa\ = 144 $\pm$ 9 \msunperyear, assuming no extinction. The \sfrpaa\ is only a small fraction of the \sfrir\ \citep[7.7 $\pm$ 0.5 \%,][]{Tan14}. We attribute this difference as due to the high extinction in GN20, especially in the central starburst and find an average extinction of A$_{V}$ = 17.2 $\pm$ 0.4 mag. Such high values are also measured in the nearby ULIRGs Arp220 and UGC 5105. 

We detect four spatially unresolved clumps in the \paa\ emission line map. By studying their \paa\ emission profile we find that each clump consists of at least two sub-clumps. These sub-clumps are also identified in the convolved F560W image. Their velocity dispersions are in the same range as measured for the CO clumps \citep{Hodge12}. The CO clumps are located more in the central region of GN20. 

We construct a \paa\ rotational profile and find it broadly consistent with the CO kinematics \citep{Hodge12}. Within the spatial resolution of the \paa\ observations (2.5 kpc) the kinematics is consistent with that of a rotationally dominated disk galaxy. This suggests that star formation is driven by gravitational instabilities in the gas rich galactic disk. The observed differences between the rotation curves could be related to the differences in spatial distribution between the CO and \paa\ emission.

We do not detect any \paa\ emission towards the UV-bright emission of GN20 as seen in HST imaging. Based on the SED fitting and the \paa\ upper limit, we conclude that this region is older than 10 Myr, where the stellar population is still bright in the UV but does not produce significant LyC radiation anymore.

Finally, we speculate on the nature of the starburst in GN20.  Evidence suggests that GN20 is in the late stage of a major merger. Based on the \paa\ kinematics we cannot distinguish between a late stage merger mimicking the disk-like rotation profile and a large extended disk, possibly re-created toward the end of the merger process.

\begin{acknowledgements}
We thank the anonymous referee for helpful and constructive comments which improved this paper.
We thank Jacqueline Hodge for providing the CO(2-1) data cube and moment maps. A.B. thanks Angela Adamo and Sean Linden for discussions on clumps and ULIRGs.
A.B., J.M. and G.\"O. acknowledge support from the Swedish National Space Administration (SNSA). L.C, J.A-M. and A.C.G. acknowledge support by grant PIB2021-127718NB-100 from the Spanish Ministry of Science and Innovation/State Agency of Research MCIN/AEI/10.13039/501100011033 and by “ERDF A way of making Europe”. F.P. and A.E. acknowledge support through the German Space Agency DLR 50OS1501 and DLR 50OS2001 from 2015 to 2023. 
K.I.C and E.I. acknowledge funding from the Netherlands Research School for Astronomy (NOVA).  K.I.C. acknowledges funding from the Dutch Research Council (NWO) through the award of the Vici Grant VI.C.212.036.
A.A.H. acknowledges support from grant PID2021-124665NB-I00  funded by MCIN/AEI/10.13039/501100011033 and by ERDF A way of making Europe.
S.G. acknowledges financial support from the Villum Young Investigator grant 37440 and 13160 and the Cosmic Dawn Center (DAWN). The Cosmic Dawn Center (DAWN) is funded by the Danish National Research Foundation under grant No. 140. T.R.G. is grateful for support from the Carlsberg Foundation via grant No. CF20-0534. 
J.P.P. and T.T. acknowledge financial support from the UK Science and Technology Facilities Council, and the UK Space Agency.
J.H and D.L. were supported by a VILLUM FONDEN Investigator grant (project number 16599)
P.G.P.-G. acknowledges support from grant PID2022-139567NB-I00 funded by Spanish Ministerio de Ciencia e Innovación MCIN/AEI/10.13039/501100011033, FEDER, UE.
 P.-O.L. acknowledges financial support from CNES
 T.R. acknowledges support from ERC grant no. 743029 (EASY).

 The work presented is the effort of the entire MIRI team and the enthusiasm within the MIRI partnership is a significant factor in its success. MIRI draws on the scientific and technical expertise of the following organisations: Ames Research Center, USA; Airbus Defence and Space, UK; CEA-Irfu, Saclay, France; Centre Spatial de Liége, Belgium; Consejo Superior de Investigaciones Científicas, Spain; Carl Zeiss Optronics, Germany; Chalmers University of Technology, Sweden; Cosmic Dawn Center (DAWN), DTU Space, Technical University of Denmark, Denmark; Dublin Institute for Advanced Studies, Ireland; European Space Agency, Netherlands; ETCA, Belgium; ETH Zurich, Switzerland; Goddard Space Flight Center, USA; Institute d'Astrophysique Spatiale, France; Instituto Nacional de Técnica Aeroespacial, Spain; Institute for Astronomy, Edinburgh, UK; Jet Propulsion Laboratory, USA; Laboratoire d'Astrophysique de Marseille (LAM), France; Leiden University, Netherlands; Lockheed Advanced Technology Center (USA); NOVA Opt-IR group at Dwingeloo, Netherlands; Northrop Grumman, USA; Max-Planck Institut für Astronomie (MPIA), Heidelberg, Germany; Laboratoire d’Etudes Spatiales et d'Instrumentation en Astrophysique (LESIA), France; Paul Scherrer Institut, Switzerland; Raytheon Vision Systems, USA; RUAG Aerospace, Switzerland; Rutherford Appleton Laboratory (RAL Space), UK; Space Telescope Science Institute, USA; Toegepast- Natuurwetenschappelijk Onderzoek (TNO-TPD), Netherlands; UK Astronomy Technology Centre, UK; University College London, UK; University of Amsterdam, Netherlands; University of Arizona, USA; University of Cardiff, UK; University of Cologne, Germany; University of Ghent; University of Groningen, Netherlands; University of Leicester, UK; University of Leuven, Belgium; University of Stockholm, Sweden; Utah State University, USA. A portion of this work was carried out at the Jet Propulsion Laboratory, California Institute of Technology, under a contract with the National Aeronautics and Space Administration. We would like to thank the following National and International Funding Agencies for their support of the MIRI development: NASA; ESA; Belgian Science Policy Office; Centre Nationale D'Etudes Spatiales (CNES); Danish National Space Centre; Deutsches Zentrum fur Luft-und Raumfahrt (DLR); Enterprise Ireland; Ministerio De Econom\'ia y Competitividad; Netherlands Research School for Astronomy (NOVA); Netherlands Organisation for Scientific Research (NWO); Science and Technology Facilities Council; Swiss Space Office; Swedish National Space Board; UK Space Agency. 

This work is based on observations made with the NASA/ESA/CSA James Webb Space Telescope. Some data were obtained from the Mikulski Archive for Space Telescopes at the Space Telescope Science Institute, which is operated by the Association of Universities for Research in Astronomy, Inc., under NASA contract NAS 5-03127 for \textit{JWST}; and from the \href{https://jwst.esac.esa.int/archive/}{European \textit{JWST} archive (e\textit{JWST})} operated by the ESDC.
\end{acknowledgements}

%
%

\bibliography{GN20_MRS.bib}
\bibliographystyle{aa}

\end{document}